\documentclass[11pt]{article}
\usepackage{graphicx,subfigure,authblk,hyperref,bbm} 
\usepackage{xcolor}
\usepackage{amsthm}
\usepackage{amssymb}
\usepackage[titletoc]{appendix}
\usepackage{mathrsfs}
\usepackage{amsfonts}
\usepackage{amsmath}
\usepackage{bm}
\usepackage[sort&compress,comma,numbers]{natbib}
\usepackage[top=2cm, bottom=2cm, outer=2cm, inner=2cm, heightrounded]{geometry}
\newcommand{\h}{\hat{H}_v}
\newcommand{\hH}{\hat{H}}
\newcommand{\f}{\mathcal{F}}
\newcommand{\dualvec}[1]{|#1\rangle}
\newcommand{\tr}{{\rm Tr}}

\newcommand{\dd }{\mathrm{d}}
\newcommand{\hvk}{{}^{\rm kin}\hat{H}_v}
\newcommand{\hek}{{}^{\rm kin}\hat{H}^E}
\newcommand{\hlk}{{}^{\rm kin}\hat{H}^L}
\newcommand{\hlv}{{\hat{H}_v^L}}
\newcommand{\hev}{{\hat{H}_v^E}}
\newcommand{\dep}{deparametrized }
\newcommand{\hh}{\hat{\bm{H}}}
\newcommand{\oo}{\mathfrak{o}}
\makeatletter
\renewcommand{\boxed}[1]{\textcolor{\boxedcolor}{%
  \fbox{\normalcolor\m@th$\displaystyle#1$}}}
  \newcommand{\makeSymbol}[1]{\mathord{\vcenter{\hbox{#1}}}}
 \newtheorem{theorem}{Theorem}[section]

\begin{document}
\title{Bouncing evolution in a model of loop quantum gravity}
\author[1,2]{Cong Zhang\thanks{zhang.cong@mail.bnu.edu.cn}}
\author[1,2]{Jerzy Lewandowski\thanks{jerzy.lewandowski@fuw.edu.pl}}
\author[1]{Haida Li\thanks{HaidaLi@mail.bnu.edu.cn}}
\author[1]{Yongge Ma\thanks{mayg@bnu.edu.cn}}
\affil[1]{ Department of Physics, Beijing Normal University, Beijing 100875, China}
\affil[2]{Faculty of Physics, University of Warsaw, Pasteura 5, 02-093 Warsaw, Poland}

\date{}
\maketitle 
\begin{abstract}
To understand the dynamics of loop quantum gravity, the deparametrized model of gravity coupled to a scalar field is studied in a simple case, where the graph underlying the spin network basis is one loop based at a single vertex. The Hamiltonian operator $\h$ is chosen to be  graph-preserving, and the matrix elements of $\h$ are explicitly worked out in a suitable basis. The non-trivial Euclidean part $\hev$ of $\h$ is studied in details. It turns out that by choosing a specific symmetrization of $\hev$, the dynamics driven by the Hamiltonian gives a picture of bouncing evolution. Our result in the model of full loop quantum gravity gives a significant echo of the well-known quantum bounce in the symmetry-reduced model of loop quantum cosmology, which indicates a closed relation between singularity resolution and quantum geometry.

\end{abstract}

\section{Introduction} 

Loop quantum gravity (LQG) is a background independent quantum theory of general relativity (GR) coupled to matter fields \cite{ashtekar2004back,han2007fundamental,rovelli2005quantum,thiemann2007modern}. The starting point of the canonical LQG is  the standard,  torsion-free Einstein's gravity in the Palatini-Holst \cite{holst1996barbero,ashtekar2004back} formulation coupled to the fields of the standard model of fundamental interactions. 
The quantization procedure begins with the reformulation of Hamiltonian GR with the Ashtekar-Barbero variables $(A_a^i,E^a_i)$, consisting of an SU(2) connection $A_a^i$ and its canonically conjugate densitized triad field $E_i^a$ on the spatial manifold $\Sigma$. 
Taking advantage of these variables, 
one can introduce the background  independent $\star$-algebra of the parallel transport (holonomy) of $A_a^i$ along all the curves and flux of $E_i^a$ (on all the 2-surfaces) subject to the canonical quantization. 
A kinematical  Hilbert space is then defined carrying quantum representation of the holonomy-flux algebra and a unitary action of the spatial diffeomorphisms as well as internal gauge transformations \cite{ashtekar1994representation,lewandowski1994topological,ASHTEKAR1995191}. 
This is a unique cyclic irreducible representation of the holonomy-flux algebra invariant with respect to semi-analytic diffeomorphisms \cite{sahlmann2006irreducibility,lewandowski2005uniqueness}. The Gaussian and vector constraints are solved exactly at quantum level and the``half-physical" space of solutions is endowed with a natural Hilbert product \cite{ashtekar1995quantization}. 
A family of operators representing geometric observables  (e.g. $2$-surface area, $3$-region volume, length of a curvature and integral of certain metric components) are regularized without need to subtract infinities and their spectra turn out to be discrete       \cite{rovelli1995discreteness,ASHTEKAR1995191,ashtekar1997quantum,ashtekar1997quantumII,thiemann1998length,ma2000qhat,ma2010new,yang2016new}.  
The Hamiltonian constraint was first rigorously regularized and promoted to an operator in \cite{thiemann1998quantum}. However, it does not preserve the diffeomorphism invariant Hilbert space. Nowadays, the dynamics of LQG is still a key open issue. Attempts to deal with this problem lead to new directions, such as the master constraint program \cite{thiemann2006phoenix,han2006master}, the algebraic quantum gravity \cite{giesel2007algebraic,giesel2010algebraic}, the deparametrized models \cite{lewandowski2011dynamics,domagala2010gravity} and the covariant LQG of spin foam models \cite{baez2000introduction,perez2003spin}.   Although much progress has been made along these approaches \cite{assanioussi2015new,
han2017einstein,assanioussi2017time}, one still need to derive dynamical predictions of LQG by means of certain specific cases or models.  

In canonical LQG, the curvature of the Ashtekar-Barbero connection emerging in the Hamiltonian constraint will result in a considerable ambiguity in 
defining a corresponding operator. 
It acts on quantum states by attaching loops to the underlying graphs. 
However, how the loops are attached is not uniquely determined. They could be defined in many different and inequivalent ways \cite{ashtekar2004back}.
A rough classification based on the ways attaching loops divides the set of all the known Hamiltonian operators of LQG into the following two categories: (i)  graph non-preserving,  (ii) graph preserving. The graph-changing action is natural from a continuum field theory approach. Several proposals of graph-changing quantum Hamiltonian operator were considered in the literatures \cite{ashtekar2004back,giesel2010algebraic,thiemann2007modern,alesci2010regularization,alesci2015hamiltonian,assanioussi2015new}. 
Self-adjointness of such operators is addressed in \cite{zhang2018towards}, where a specific graph-changing Hamiltonian operator is proven to be self-adjoint in a domain of certain special states. 
The second category, that is the graph persevering action, is the one considered in the current paper.  The action of such an operator is reducible to subspaces corresponding to the graphs of given states. For a given graph,  the analytic properties of the operator are much easier to study than in general cases. In the model of GR deparametrized by a dust field, it is argued that the graph-preserving action of the Hamiltonian operator is the only diffeomorphism-invariant option \cite{giesel2010algebraic}.

Applying the idea of loop quantization to symmetry-reduced models of cosmology has resulted in an active field called loop quantum cosmology (LQC) \cite{ashtekar2003mathematical}. 
Notable progress has been made in the quantum cosmological model,
 especially the classical big-bang singularity resolution which was first discussed in \cite{bojowald2001absence} by considering the discrete feature of the model and later was realized in some improved treatment  by the quantum bounce scenario in  \cite{ashtekar2006quantumnature,ding2009effective,yang2009alternative}.
However, the symmetry reduction in LQC is done before quantization. Recently, attempts of defining dynamics for LQG and using that to generate a model of quantum cosmology are investigated, for examples, the spinfoam cosmology \cite{bianchi2010towards}; the quantum reduced loop gravity approach \cite{alesci2015loop}; the arising of LQC dynamics as
the action of the full theory Hamiltonian on maximally coarse states in the kernel of suitably chosen reduction
constraints \cite{bodendorfer2016embedding}; the group field theory condensates approach to quantum cosmology \cite{gielen2016quantum};
calculation of the expectation value of the scalar constraint with respect to some coherent states peaked on the phase-space variables of flat Robertson–Walker spacetime in \cite{dapor2018cosmological},  which reproduces the effective Hamiltonian constraint derived in \cite{yang2009alternative} by an alternative quantization of the Hamiltonian in LQC. 
However, how LQG solves the prevailing singularities in GR in full theory is still unclear. It is crucial to check whether a similar picture as the quantum bounce resolution of singularity in LQC can be achieved in full LQG. 

In the current paper, the full LQG dynamics is studied by focusing on a specific case of quantum states with particular graphs. Rather than the symmetry-reduced model of LQC in \cite{ashtekar2006quantumnature}, our model contains the full degree of freedom. The model will be studied in the theory of GR coupled to a massless scalar field. 
It is one of the two known  remarkable cases  in which the Dirac quantization program can be completed  by using the deparametrized procedure \cite{domagala2010gravity,lewandowski2011dynamics}. 
In this theory, several results were derived. First,
all the quantum constraints of the canonical GR were solved completely 
and a general solution was given explicitly, by assuming the existence of certain operators. Thus, the physical Hilbert space of the solutions was defined.  
Second, the general formula for Dirac observables that commutes with all the constraints  was derived.  
Finally, the resulting algebra of the Dirac observables  was shown to admit an action of the $1$-dimensional group of automorphisms  that classically corresponds to the transformations of adding  a constant to the scalar field.  
The generator of those automorphisms was promoted to the physical quantum Hamiltonian operator $\hat{H}_{\rm phys}$ of the system. 
An exact derivation of that operator in LQG has become possible \cite{domagala2010gravity,lewandowski2011dynamics}. 
Different from the previous works, in the model considered in the current paper the Hamiltonian operator can be well-defined directly in the graph-preserving category without  introducing some extra Hilbert spaces.

Due to the reducibility  of the Hamiltonian operator to subspaces corresponding to the graphs, we can focus on a specific graph, which in the current paper is chosen to be a loop based at a single vertex. In this case, the action of the Hamiltonian operator changes only the spins on these the loop. The matrix elements of the Hamiltonian operator on the spin network basis are calculated. Using the asymptotic behaviour of the matrix elements, we can prove the self-adjointness of the operator based on our previous work \cite{zhang2018towards}. Then the operator is diagonalized by spectral decomposition technique, and the dynamical evolution  of some specific initial coherent state driven by the Euclidean part of the Hamiltonian is considered. Our analysis of the dynamics reveals a quantum bounce picture of the evolution similar to that in LQC. 

The paper is organized as follows. In Sec. \ref{se:two}  the \dep model of LQG coupled to a scalar field is  briefly reviewed. In Sec. \ref{se:three}, we restrict ourself to a simple case and derivative the matrix elements of the Hamiltonian operator by choosing a suitable basis. In Sec. \ref{se:four}, we show that the spectrum of the Euclidean part $\hev$ of the Hamiltonian operator is the entire real line. Then the dynamics driven by the Euclidean part is studied and a picture of quantum bounce evolution is obtained in Sec. \ref{se:dynamic}. Finally, the present work is concluded and discussed in Sec. \ref{se:conclution}.

\section{A general work on \dep model}\label{se:two}
\subsection{The classical theory}
Considering gravity minimally coupled to a massless Klein-Gordon field in the ADM formalism with Ashtekar-Barbero variables, we have a totally constrained system with the standard canonical variables $(A^i_a(x),E^a_i(x))$ for gravity and $(T(x),\pi(x))$ for scalar field defined at every point $x$
 of an underlying $3$-dimensional manifold $\Sigma$. The diffeomorphism and scalar constraints are respectively
\begin{equation}\label{eq:vectorconstraint}
C_a(x)=C_a^{\rm gr}(x)+\pi(x)T_{,a}(x)=0,
\end{equation}
\begin{equation}\label{eq:scalarconstraint}
C(x)=C^{\rm gr}(x)+\frac{1}{2}\frac{\pi(x)^2}{\sqrt{|\det E(x)}|}+\frac{1}{2}q^{ab}(x)T_{,a}(x)T_{,b}(x)\sqrt{|\det E(x)|}=0,
\end{equation}
where $C^{\rm gr}_a$ and $C^{\rm gr}$ are the vacuum gravity constraints and $q^{ab}=\frac{E^a_iE^b_i}{|\det E|}$. 

The \dep procedure starts with assuming that the constraints \eqref{eq:vectorconstraint} are satisfied. By replacing $T_{,a}$  by $-C_a^{\rm gr}/\pi$, the constraints \eqref{eq:scalarconstraint} are rewritten as
\begin{equation}
\pi^2=\sqrt{|\det E|}\left(-C^{\rm gr}\pm \sqrt{(C^{\rm gr})^2-q^{ab}C^{\rm gr}_aC^{\rm gr}_b}\right).
\end{equation}
 The sign ambiguity is solved depending on a quarter of the phase space. We choose the one that contains the homogeneous cosmological solutions \cite{alesci2015hamiltonian}.  
 In that part of the phase space, the scalar constraint $C(x)$ can be replaced by,  
 \begin{equation}\label{eq:newfour}
C'(x)=\pi(x)\pm \sqrt{h(x)},
\end{equation}
where
\begin{equation}\label{eq:classicalhamil}
h=\sqrt{|\det E|}\left(-C^{\rm gr}+ \sqrt{(C^{\rm gr})^2-q^{ab}C^{\rm gr}_aC^{\rm gr}_b}\right).
\end{equation}
\subsection{The structures of the quantum theory }
For the \dep  theory, the entire Dirac quantization scheme can be implemented and performed without major setbacks \cite{domagala2010gravity,lewandowski2011dynamics}.  The result is a physical Hilbert space of solutions to the constraints, together with algebra of quantum Dirac observables endowed with one dimensional group of automorphisms generated by a quantum Hamiltonian operator. This resulting structure is equivalent to  the following 
model that is expressed in a derivable way by elements of the framework of LQG.
\begin{itemize}
\item The physical Hilbert space $\mathcal{H}$ is the space of the quantum states of the vacuum (matter free) gravity
in terms of the Ashtekar-Barbero connection-frame variables that satisfy the vacuum quantum vector constraint and the vacuum quantum Gauss constraint.  
In other words, in the connection representation,  the states are constructed from functions $A\mapsto \Psi(A)$ invariant with respect to the 
diffeomorphism transformations
$$ A' = f^*A, \ \ \ \ \ \ \ \ \ \ \ \ \forall f\in {\rm Diff}(\Sigma)$$
and to the Yang-Mills gauge transformations 
\begin{equation} \label{YM}
 A' = g^{-1}Ag + g^{-1} dg, \ \ \ \ \ \ \ \ \ \ \ \ \forall g\in C(\Sigma,G).
  \end{equation}
They are not assumed to satisfy the vacuum scalar constraint, though. Such Hilbert space is available in the LQG framework.  

\item The Dirac observables are represented by the set of operators $\{\hat{\mathcal{O}}\}$  in $\mathcal{H}$. When the scalar field transforms as $T\mapsto T+t$ with a constant $t$, the observables transform as
\begin{equation}\label{eq:evoop}
\hat{\mathcal{O}}\mapsto e^{i\hH t}\hat{\mathcal{O}} e^{-i\hH t}
\end{equation}
Therefore the quantum dynamics in the Schr\"odinger picture  is given by
\begin{equation}
 i\frac{d}{dt}\Psi=\hH\Psi,
 \end{equation} 
 $\hH$ is called the quantum Hamiltonian.

 \item The quantum Hamiltonian 
 \begin{equation}\label{eq:integralh}
 \hH=\int d^3x \widehat{\sqrt{-2\sqrt{|\det E(x)|}C^{\rm gr}(x)}}
 \end{equation}
 is a quantum operator corresponding to the classical physical Hamiltonian $$H=\int d^3x \sqrt{-2\sqrt{|\det E(x)|}C^{\rm gr}(x)},$$ 
where the term $q^{ab}C_a^{\rm gr}C_b^{\rm gr}$ in \eqref{eq:classicalhamil} is dropped because $\hat{H}$ will finally be defined on the diffeomorphism invariant states that are supposed to be in the kernel of the operator $\hat{C}_a^{\rm gr}$ representing $C_a^{\rm gr}$ and we choose the operators ordering such that $\hat{C}_a^{\rm gr}$ is always to the right in the expansion of the  square-roots in \eqref{eq:newfour}.
\end{itemize} 
 
The classical Hamiltonian $H$  is manifestly spatial diffeomorphism invariant, and the same is expected of the quantum Hamiltonian operator 
$\hH$. There seems to be a perfect compatibility between the diffeomorphism invariance of the quantum Hamiltonian operator and the diffeomorphism invariance of the quantum states, elements of the Hilbert space ${\cal H}$.  
However, the integrand $\sqrt{-2\sqrt{|\det E(x)|}C^{\rm gr}(x)}$ involves the square root of an expression assigned to each point $x$. 
In order to quantize this integrand, the operator corresponding to the expression of $-2\sqrt{|\det E(x)|}C^{\rm gr}(x)$  under the square root should be obtained first. However, $-2\sqrt{|\det E(x)|}C^{\rm gr}(x)$ alone is not diffeomorphism invariant, which leads to the fact that the corresponding operator can not be well defined in the diffeomorphism invariant Hilbert space. 
The kinematical Hilbert space ${\cal H}_{\rm kin}$ can be chosen to define the operator $\widehat{-2\sqrt{|\det E(x)|}C^{\rm gr}(x)}$ corresponding to $-2\sqrt{|\det E(x)|}C^{\rm gr}(x)$. Only if $\widehat{-2\sqrt{|\det E(x)|}C^{\rm gr}(x)}$ on   ${\cal H}_{\rm kin}$ is self-adjoint and non-negative, its square root, as well as the operator $\hat{H}$ in \eqref{eq:integralh}, is well defined in ${\cal H}_{\rm kin}$. Solving the diffeomorphism constraints lead to the physical Hilbert space $\mathcal{H}$ which is a dual space of a dense subspace of $\mathcal{H}_{\rm kin}$. The operator $\hat{H}$ then can be passed to the Hilbert space $\mathcal{H}$ naturally by the dual action since it is diffeomorphism invariant.
   
\subsection{The Hilbert spaces  ${\cal H}_{\rm kin}$ and ${\cal H}$} 
The kinematical Hilbert space ${\cal H}_{\rm kin}$ of the vacuum LQG  consists of functions 
\begin{equation}\label{cyl}
\Psi_\gamma(A)\ =\ \psi_\gamma(h_{e_1}(A), ... , h_{e_n}(A)),
\end{equation}
where $e_1,...,e_n$ are the edges of a finite graph $\gamma$ embedded in $\Sigma$, and  $h_{e}(A)\in SU(2)$ is the parallel transport along 
a path $e$ in $\Sigma$ with respect to a given connection $1$-form $A$,
$$ h_e(A)\ =\mathcal{P} \exp(- \int_e A). $$
  In the LQG framework those functions of the variable $A$ are called cylindrical functions.
It can be also defined as define a multiplication operator, given a representation $D^{(l)}$ of SU(2), 
$$ (D^{(l)}{}^m{}_n (h_e(A))) \Psi(A)\ =\  D^{(l)}{}^m{}_n (h_e(A))\Psi(A),$$
where $m,n$ label an entry of the representation matrix. The kinematical space can be decomposed  into the orthogonal sum 
\begin{equation}\label{decompkin} 
{\cal H}_{\rm kin}\ =\ \overline{\bigoplus_\gamma {\cal H}_\gamma},
\end{equation}
where $\gamma$ runs through the set of embedded finite graphs in $\Sigma$ (un-oriented). We also use a basis $\tau_1,\tau_2,\tau_3 \in{\rm su}(2)$ such that
$$ [\tau_i,\tau_j]\ =\ \epsilon_{ijk}\tau_k . $$ 
Another operator employed in the current paper is the `angular momentum operator' $J^i$ defined in ${\cal H_\gamma}$. Given a graph $\gamma$, a vertex $v$, 
and an edge $e_0$  at $v$, and  $\tau_i$, it acts  on the function  $\Psi_\gamma$ of \eqref{cyl} as \cite{ashtekar2004back}
\begin{equation}\label{J}
(J_{v,e_0}^i\Psi_\gamma)(A)=\left\{
\begin{matrix}
i \left.\frac{d}{dt}\right|_{t=0}\psi_\gamma(\cdots,h_e(A),e^{ -t\tau_i}h_{e_0}(A),g_{e'}(A),\cdots),& v=t(e_0)\\
i \left.\frac{d}{dt}\right|_{t=0}\psi_\gamma(\cdots,h_e(A),h_{e_0}(A)e^{t\tau_i},h_{e'}(A),\cdots),& v=s(e_0).
\end{matrix}
\right.
\end{equation}
where $t(e)$ and $s(e)$ represent the target and source of the edge $e$\footnote{In order to define the operator, the graph must be oriented. However, physical states are defined by the equivalence classes modulo the orientation \cite{thiemann2007modern}. }. 

In this paper we restrict to  the  functions  invariant with respect to  the Yang-Mills gauge transformations (\ref{YM}).  
An orthonormal basis can be constructed from the spin-network states. 
Given a graph $\gamma$ in $\Sigma$, we denote by $V(\gamma)$ the set of the vertices and $E(\gamma)$ the set of the edges. 
The symmetries of $\gamma$ are denoted by ${\rm Diff^{s\omega}_\gamma}$ with $s\omega$ representing that the diffeomorphism  is semi-analytic \cite{lewandowski2005uniqueness} and the elements of ${\rm Diff}^{s\omega}_\gamma$ preserving every edge of $\gamma$  by ${\rm TDiff}(\gamma)$. We then denote $\mathcal{S}^{s\omega}_\gamma:={\rm Diff}_\gamma^{s\omega}/{\rm TDiff}(\gamma)$.
Given any cylindrical function $\Psi_\gamma\in{\cal H}_\gamma$, one proceeds to two steps to get the corresponding diffeomorphism invariant state. First,
$\psi_\gamma$ averaged by using only the symmetries of $\gamma$ and an projection $\eta$ is defined as
\begin{equation}\label{eta}
\eta:\Psi_\gamma\mapsto \frac{1}{\#(\mathcal{S}^{s\omega}_\gamma)} \sum_{f\in \mathcal{S}_\gamma^{s\omega}}   U_f\cdot \Psi_\gamma 
\end{equation}
 where $ U_f$ denotes the unitary operator corresponding to the diffeomorphism transformation $f$ on $\Sigma$ and $\#(\mathcal{S}^{s\omega}_\gamma)$ is the order of the group $\mathcal{S}^{s\omega}_\gamma$ \cite{ashtekar2004back}.
Then it is expected to perform the group-averaging with respect to the remaining diffeomorphisms which move the graph $\gamma$ and define a map $P_{\rm diff}$ as
\begin{equation}\label{eq:P}
 P_{\rm diff}:\Psi_\gamma\mapsto\sum_{f\in {\rm Diff}^{s\omega}/{\rm Diff}^{s\omega}_\gamma} \langle U_f\cdot \eta\Psi_\gamma|.
 \end{equation} 
 Thanks to the projection $\eta$ in \eqref{eq:P}, we can restrict $P_{\rm diff}$ to the subspace $\eta(\mathcal{H}_\gamma)$ such that $P_{\rm diff}$ maps elements in $\eta(\mathcal{H}_\gamma)$  into the algebraic dual $\left(\bigoplus_\gamma \eta({\cal H}_\gamma)\right)'$. In other words, $P_{\rm diff}(\mathcal{H}_\gamma)=P_{\rm diff}(\eta(\mathcal{H}_\gamma))$. Diffeomorphism invariant operators in $\mathcal{H}_\gamma$ have to preserve $\eta(\mathcal{H}_\gamma)$ and can be defined in $P_{\rm diff}(\mathcal{H}_{\gamma})$ naturally by duality. Therefore without losing the generality, we will restrict ourself to the subspace $\mathcal{H}_\gamma^s:=\overline{\eta(\mathcal{H}_\gamma)}\subset \mathcal{H}_\gamma$. Let $[\gamma]$ be the equivalence class of the graphs diffeomorphism-equivalent to $\gamma$. Then we fix once for all a representative from each equivalence class, collect all of the representatives to be a set $\Gamma_{\rm diff}$ and define 
 \begin{equation}
 \mathcal{H}_{\rm kin}^s:=\bigoplus_{\gamma\in \Gamma_{\rm diff}}\mathcal{H}_\gamma^s.
 \end{equation}
 It is easy to conclude that $\mathcal{H}_{\rm kin}^s$ is a subspace of the whole kinematical Hilbert space $\mathcal{H}_{\rm kin}$ and, instead of $\mathcal{H}_{\rm kin}$,
it is sufficient to consider the space $\mathcal{H}_{\rm kin}^s$ itself.

\subsection{The physical quantum Hamiltonian operator}\label{se:phyH} 
In the current paper we will consider the general regularization scheme for 
the operator $\widehat{\sqrt{-2\sqrt{|\det E(x)|}C^{\rm gr}(x)}}$  introduced in \cite{alesci2015hamiltonian,yang2015new}. 
However, instead of adding any new edge to the graph $\gamma$ by the operator $\hek_{v,ee'}$ as defined in \cite{zhang2018towards}, we use the loops constituting $\gamma$ to regulate the curvature and ignore the limit process in the classical expression.
According to the framework, operator $\widehat{\sqrt{-2\sqrt{|\det E(x)|}C^{\rm gr}(x)}}$ is defined on $\mathcal{H}_{\rm kin}$ as
\begin{equation}\label{theoperator}
\widehat{\sqrt{-2\sqrt{|\det E(x)|}C^{\rm gr}(x)}}\ =\ \sum_{v\in \Sigma}\delta(v,x)\sqrt{\hvk}
\end{equation}
where $\delta(v,x)$ is the Dirac distribution and $\hvk$ is a well-defined operator acting only on vertices of a graph.
The sum seems to be awfully infinite. However, for every subspace $ {\cal H}_\gamma$, the only non-zero
terms correspond to the vertices of the underlying graph $\gamma$ of a cylindrical function.
For every vertex $v\in V (\gamma)$  the operator ${}^{\rm kin}\h$ is defined first  as an operator
in the kinematical Hilbert subspace ${\cal H}_\gamma$ and then passed to be an operator in $\mathcal{H}_\gamma^s$. A subtle issue here is the self-adjointness and the positive definiteness of ${}^{\rm kin}\h$. To implement these two properties, we will consider the operator $\sqrt{\frac{1}{2}\left({}^{\rm kin}\h({}^{\rm kin}\h)^\dagger+({}^{\rm kin}\h)^\dagger{}^{\rm kin}\h\right)}$.

 As in \cite{alesci2015hamiltonian,assanioussi2015new,yang2015new,zhang2018towards} $\hvk$  takes the form  of the sum with respect to the pairs of edges $(e,e')$ at $v$,
\begin{equation*}
\hvk = \sum_{e,e'\text{ at } v}\epsilon(e,e'){}^{\rm kin}\hat{H}_{v,e,e'}
\end{equation*}
where $\epsilon(e,e')$ equals to $0$ if the edges  $e$ and $e'$ are tangent at $v$ or $1$ otherwise. However, in the current paper, the graph-preserving Hamiltonian is considered. Then, $\hvk$ can be rewritten as 
\begin{equation}\label{eq:physicalH}
\hvk = \sum_{\alpha\text{ at } v}\epsilon(\alpha){}^{\rm kin}\hat{H}_{v,\alpha}
\end{equation}
where $\epsilon(\alpha)$ equals to $0$ if the loop $\alpha$ based at $v$ is differentiable at $v$ or $1$ otherwise, and for a given graph the minimal loop \cite{giesel2007algebraic} $\alpha$ contained in the graph at each vertex. 
The operator  at each minimal loop of a vertex  consists of the so-called Lorentzian and Euclidean parts,
$${}^{\rm kin}\hat{H}_{v,\alpha}\ =\ (1+\beta^2)\hlk_{v,\alpha}+ \hek_{v,\alpha},
$$ 
where  the operators $\hek_{v,\alpha}$ and $\hlk_{v,\alpha}$ act  on a cylindrical function as follows
\begin{itemize}
\item Denote $\alpha_b$ and $\alpha_e$ the beginning and the ending segments of the loop $\alpha$ respectively. By using \eqref{J}, we have two operators $J_{v,\alpha_b}$ and $J_{v,\alpha_e}$. Then $\hek_{v,\alpha}$ is defined as
\begin{equation}\label{eq:eucleadianH}
\hek_{v,\alpha}\ =\ \kappa_1\epsilon_{ijk} J^j_{v,\alpha_e}J^k_{v,\alpha_b}(h^i_{\alpha})^{(l)},
\end{equation}    
where
\begin{equation}
\begin{aligned} 
(h^i_{\alpha })^{(l)} \ &:=\ -\frac{3}{l(l+1)(2l+l)}\ {\rm Tr} \left(D^{(l)}(h_{\alpha }) D'^{(l)}(\tau^i)\right).
\end{aligned}
\end{equation}
Here the factor $\kappa_1$ is arbitrary, representing a residual ambiguity of the quantization which will be setted as 1 in the present work for simplicity. The irreducible  representation $l$ 
of SU(2) on the minimal loop $\alpha$  is chosen such that the spin network decomposition of the resulted cylindrical function does not contain a zero spin at any segment of $\alpha$. In the simple model considered in next section, $l$ can be fixed as $1/2$ by restricting the domain of the Hamiltonian operator.
 \item  $\hlk_{v,\alpha}$ is given directly by
 \begin{equation}\label{eq:lor}
 \begin{aligned}
\hlk_{\alpha,v}:=&\sqrt{\delta^{ii'}\left(\epsilon_{ijk}J_{v,\alpha_b}^jJ_{v,\alpha_b}^k\right)\left(\epsilon_{i'j'k'}J_{v,\alpha_e}^{j'}J_{v,\alpha_e}^{k'} \right)}\times\\
&\left(\frac{2\pi}{\mathfrak{A}}-\pi+\arccos\left[\frac{\delta_{kl}J_{v,\alpha_b}^kJ_{v,\alpha_e}^l}{\sqrt{\delta_{kk'}J_{v,\alpha_b}^kJ_{v,\alpha_b}^{k'}}\sqrt{\delta_{kk'}J_{v,\alpha_e}^kJ_{v,\alpha_e}^{k'}}}\right]\right),
\end{aligned}
\end{equation}
where the factor $\mathfrak{A}$ is arbitrary, representing another residual ambiguity of the quantization. 
\end{itemize}

In order to implement (\ref{theoperator}), one has  to find a basis in ${\cal H}_{\rm kin}^s$ that consists of eigenstates of  $\hvk$ satisfying 
$$   \hvk |v,\lambda\rangle=\lambda |v,\lambda\rangle.$$
We are interested in an operator whose action on the above eigenstates reads 
\begin{equation}\label{theoperator'}
\widehat{\sqrt{-2\sqrt{|\det E(x)|}C^{\rm gr}(x)}}\,|v,\lambda\rangle\ =\ \sum_{v\in \Sigma}\delta(v,x)\sqrt{\lambda}\,|v,\lambda\rangle .
\end{equation}
For the time being,  we do not even know  a single non-trivial eigenstate of the operator $\hvk$. In the next section we will restrict our study in the simplest  subspace of ${\cal H}_{\rm kin}^s$, which contains states of  a loop based at a single vertex . The properties of the operator $\hvk$ are studied therein.

\section{The simple case of one loop}\label{se:three}
  In this section, we consider the simplest case where the graph $\gamma$ contains only one  loop $\alpha$ based at a vertex $v$.
The graph $\gamma$ defines a kinematical subspace $\mathcal{H}_\gamma$ and symmetrized space $\mathcal{H}_\gamma^s$. Let $\mathcal{H}_\gamma^G\subset\mathcal{H}_\gamma$ be the gauge invariant subspace.
A spin network basis, denoted as $|j\rangle$, of the Hilbert space $\mathcal{H}^G_\gamma$ is defined as 
\begin{equation}\label{eq:basis}
\langle A|j\rangle:={\rm Tr}(D^j(h_\alpha( A)),
\end{equation}
where $D^j$ on the right hand side is the Wigner-D matrix  of the holonomy $h_\alpha(A)$.

The next step in the construction is to consider the projection $\eta$ in \eqref{eta}.  The graph $\gamma$ has the symmetries $\mathcal{S}_\gamma$ generated by  homeomorphism of $f$ given by
\begin{equation}
\begin{aligned}
&f(\alpha)=\alpha^{-1}
\end{aligned}
\end{equation}
By definition, we have 
\begin{equation}
\begin{aligned}
\langle A |U_{f}|j\rangle=\tr (D^j(h_{\alpha}(A)^{-1})=\tr(D^j(h_{\alpha}(A)), 
\end{aligned}
\end{equation}
and hence $$U_{f}|j\rangle= |j\rangle.$$ 
Thus we have 
\begin{equation}\label{eq:symmtrization}
\frac{1}{\#(\mathcal{S}_\gamma^{s\omega})}\sum_{f\in\mathcal{S}^{s\omega}_\gamma}U_f|j\rangle=|j\rangle,
\end{equation}
which leads to 
\begin{equation}
\mathcal{H}_\gamma^s=\mathcal{H}_\gamma.
\end{equation}
\subsection{The action of $H_v$ on $\mathcal{H}_v$}
We now calculate the action of the operator $\h$ defined by \eqref{eq:physicalH} on the Hilbert space $\mathcal{H}_\gamma$. For the simple graph we consider, it has
\begin{equation}
\hvk=\hek_{v,\alpha}+(1+\beta^2)\hlk_{v,\alpha}.
\end{equation}
The equation \eqref{eq:lor} implies that the Lorentz part $\hlv:=\hlk_{v,\alpha}$ can be diagonalized in our basis as
\begin{equation}\label{eq:lorentzpart}
\begin{aligned}
\hlv|j\rangle=\sqrt{j(j+1)}|j\rangle=:h_l(j)|j\rangle.
\end{aligned}
\end{equation}

In the rest of the paper, we will focus on the properties of the Euclidean part rather than the Lorentzian one.
The former is more complicated than the latter. As shown in appendix \ref{app:euaction}, for $j\neq 0$ we obtain
\begin{equation}
\begin{aligned}
&\hek_{v,\alpha}| j\rangle=(j+\frac{3}{2})|j+1/2\rangle-(j-\frac{1}{2})|j-1/2\rangle.
\end{aligned}
\end{equation}
For $j=0$, we assume that the factor $\tr(h_\alpha\tau_i)$ of $\hek_{\alpha,v}$ creates a differentiable loop $\alpha$. Hence we get $\hek_{\alpha,v}|j=0\rangle=0$ because of the factor $\epsilon(\alpha)$ in \eqref{eq:physicalH}. 
 It should be noticed that in the case of  $j=1/2$, one has
\begin{equation}
\begin{aligned}
&\hek_{v,\alpha}| 1/2\rangle=2|1\rangle.
\end{aligned}
\end{equation}
Then by definition, the action of the adjoint operator $(\hek_{v,\alpha})^\dagger$ reads
\begin{equation}
\begin{aligned}
&(\hek_{v,\alpha})^\dagger| j\rangle=-j|j+1/2\rangle+\Xi(j)(j+1)|j-1/2\rangle\\
&(\hek_{v,\alpha})^\dagger| j=0\rangle=0
\end{aligned}
\end{equation}
where $\Xi(j):=1-\delta_{j,1/2}$ is equal to $0$ if $j=1/2$ and $1$ otherwise.  A symmetric Hamiltonian operator $\hev$ is then defined by
  $$\hev:=\sqrt{\frac{1}{2}\Big(\hek_{v,\alpha}(\hek_{v,\alpha})^\dagger+(\hek_{v,\alpha})^\dagger(\hek_{v,\alpha})\Big)}=:\sqrt{\hh}.$$ 
It is easy to see that
\begin{equation}
\hh|0\rangle=0,
\end{equation}
and for $j\neq 0$
  \begin{equation}\label{eq:hh}
\begin{aligned}
&\hh|j\rangle=C_+(j)\dualvec{j+1}\rangle+C_0(j)\dualvec{j}+C_-(j)\dualvec{j-1}.
\end{aligned}
\end{equation}
with
$$
C_+(j)=-(j^2+2j+\frac{3}{8}),~C_-(j)=\Xi(j)\Xi(j-1/2)C_+(j-1),~C_0(j)=j^2+\frac{3}{2}j+\frac{9}{8}+\Xi(j)(j^2+\frac{1}{2}j+\frac{5}{8}).
$$
One can show that $C_0(j)=-C_+(j)-C_-(j)+2$ for $j>1$. 
The operator $\hev$ and $\hh$ are unbounded and thus can not be defined on the whole $\mathcal{H}_\gamma^s$. A natural choice of the domain for both of them is
 \begin{equation}
  \f=\left\{\dualvec{\psi}:\langle j \dualvec{\psi}\neq 0 \text{ for finite number of } |j\rangle\right\}.
  \end{equation} 
  It is obvious that  $\f$ consist of smooth cylindrical functions. 
The operators $\hev$ and $\hh$  with the domain $\f$ are symmetric, densely defined in $\mathcal{H}_\gamma^s$. Hence they admits closures. We will refer to these operators as their closures in the following sections.
\section{Spectrum of $\hev$ }\label{se:four}
 Let us define
\begin{equation}\label{eq:HRI}
\hat{H}_R:=\frac{1}{2}(\hek_{v,\alpha}+(\hek_{v,\alpha})^\dagger),~\hat{H}_I:=\frac{-i}{2}(\hek_{v,\alpha}-(\hek_{v,\alpha})^\dagger).
\end{equation}
Then both of $\hat{H}_R$ and $\hat{H}_I$ are symmetric operators defined on $\f$. By definition, for $j\neq 0$ it is to obtain
\begin{equation}
\hat{H}_R\dualvec{j}=\frac{3}{4}\dualvec{j+1/2}+\Xi(j)\frac{3}{4}\dualvec{j-1/2},
\end{equation}
and
\begin{equation}\label{eq:halphonbasis}
\hat{H}_I\dualvec{j}=-i(j+\frac{3}{4})\dualvec{j+1/2}+i\Xi(j)(j+\frac{1}{4})\dualvec{j-1/2}.
\end{equation}
Thus
$\hat{H}_R$ is obviously self-adjoint since it is bounded with $\|\hat{H}_R\|\leq 3/2$. For the operator $\hat{H}_I$, we conclude from \eqref{eq:halphonbasis} that
\begin{itemize}
\item each state $|j\rangle$ is mapped by $\hat{H}_I$ into a linear combination of finite  elements of that basis;
\item the coefficients of the terms in the combination depend on $j$ linearly.
\end{itemize} 
Thus by the same technique as in \citep[Lemma 4.0.1]{zhang2018towards}, one can find some operator $\hat{N}>1$, with which there exist $c,d\in\mathbb{R}^+$ such that for all $|\psi\rangle\in\f$ the following conditions are satisfied:
\begin{equation*}
\begin{aligned}
\|\hat{H}_I|\psi\rangle\|^2&\leq c\|\hat{N}|\psi\rangle\|^2\\
\left| \langle\psi|[\hat{H}_I,\hat{N}]|\psi\rangle\right|&\leq d\|\sqrt{\hat{N}}|\psi\rangle\|^2.
\end{aligned}
\end{equation*}
These equations guarantee the self-adjointness of $\hat{H}_I$. Therefore, $\hat{H}_R^2$ and $\hat{H}_I^2$ are also self-adjoint. Since $\hat{H}_R$ is bounded while $\hat{H}_I$ is unbounded, it is concluded that $\hh=H^2_I+H^2_R$ is self-adjoint \citep{reed1975}. 

By definition, the state $|0\rangle$ is an eigenstate of the operators $\hev$ and $\hh$, with the eigenvalue $0$. 
If we consider a state $|\psi\rangle=\sum_{j=1/2}^\infty\psi_j|j\rangle$. By using \eqref{eq:hh}, we have
\begin{equation}
\begin{aligned}
&\langle\psi|\hh|\psi\rangle
\geq \sum_{j=1/2}^\infty C_0(j)|\psi_j|^2-|C_+(j)|\frac{|\psi_{j+1}|^2+|\psi_j|^2}{2}-|C_-(j)|\frac{|\psi_{j-1}|^2+|\psi_j|^2}{2}
\geq \frac{1}{2}\langle\psi|\psi\rangle
\end{aligned}
\end{equation}
which leads to that $\sigma(\hh)-\{0\}\geq 1/2$ with $\sigma(\hh)$ denoting the spectrum of $\hh$. We will show below that $\sigma(\hat{H}_I)=\mathbb{R}$ and accordingly $\sigma(\hat{H}_I^2)=\mathbb{R}^+$. Combining these results, one can get that $\sigma(\hh)=\{0\}\cup[\mathfrak{E},\infty)$ with some $\mathfrak{E}\geq 1/2$. 

\subsection{The spectrum for $\hat{H}_I$}\label{se:4p1}
\begin{figure}
\begin{center}
\includegraphics[width=0.8\textwidth]{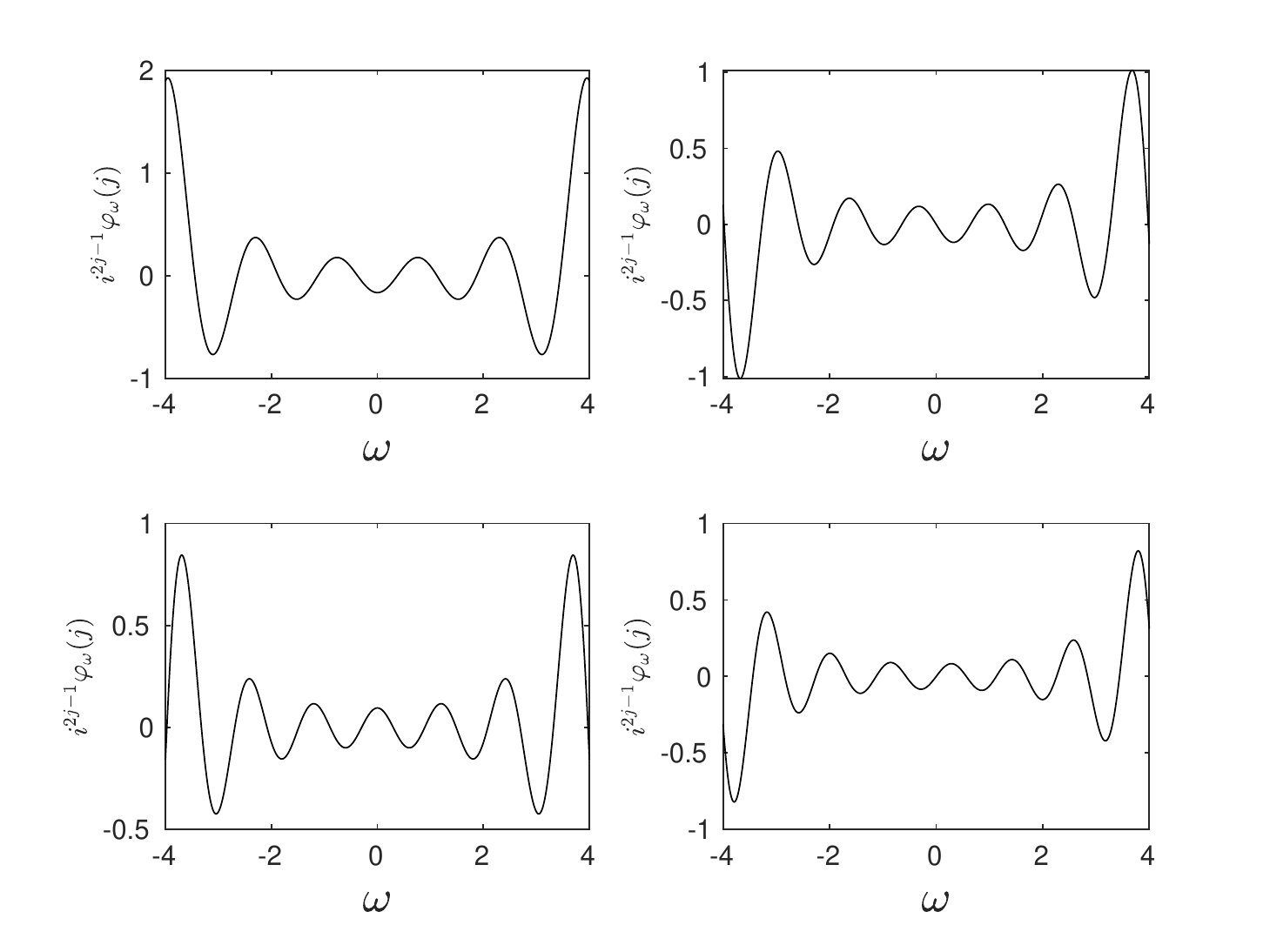}
\end{center}
\caption{Behaviours of $\varphi_\omega(j)$ as a function of $\omega$ for various spin $j$. The values of $j$ from the up left and right to the down left and right are $j=75/2$, $j=75$, $j=225/2$ and $j=150$ respectively. As shown in the plots, when $\omega$ is far from the maximal and minimal roots  of the polynomial $\varphi_\omega(j)$, $\varphi_\omega(j)$ oscilates.}\label{fig:plotomega1}
\end{figure}

 Let $|\omega\rangle=\sum_j\varphi_\omega(j)|j\rangle$ be such a state that $
\hat{H}_I|\omega\rangle=\omega|\omega\rangle$. By definition, if $\omega=0$, there are two kinds of solutions
\begin{itemize}
\item[(i)] $\varphi_{\omega=0}(0)=1$ and $\varphi_{\omega=0}(j)=0,~\forall j\neq 0$;
\item[(ii)] $\varphi_{\omega=0}(0)=0$ and $\varphi_{\omega=0}(j)\neq 0,~\forall j\neq 0$.
\end{itemize}
 However, for the case $\omega\neq 0$, only the second kind of the above solutions is available. Therefore, without lose of generality, we will consider the case $|\omega\rangle=\sum_{j\neq 0}\varphi_\omega(j)|j\rangle$. 
According to \eqref{eq:halphonbasis}, when $j\neq 0$ the eigen-equation reads as
\begin{equation}\label{eq:recurrence}
 \begin{aligned}
\omega\varphi_\omega(j)&=i\Xi(j)(j+\frac{1}{4})\varphi_{\omega}(j-1/2)-i(j+\frac{3}{4})\varphi_{\omega}(j+1/2).
\end{aligned}
\end{equation}
Defining
$$
 \xi(x):=\sqrt{x^2+\frac{1}{16}} 
$$
we can rewrite \eqref{eq:recurrence} as
\begin{equation}\label{eq:simplerecurrence}
\omega\varphi_\omega(j)=i\Xi(j)\xi(\sqrt{j(j+1/2)}~)\varphi_\omega(j-1/2)+i\xi(\sqrt{(j+1)(j+1/2)}~)\varphi_\omega(j+1/2).
\end{equation}
Given the initial condition:  $\varphi_\omega(1/2)=1~\forall \omega$, the solution to \eqref{eq:simplerecurrence} takes the form
\begin{equation}\label{eq:varphiformula}
\varphi_\omega(j)=i^{2j-1}\sum_{\mu=0}^{\lfloor j-\frac{1}{2}\rfloor}\zeta_{j,\mu}\omega^{2j-1-2\mu},
\end{equation}
where $\lfloor x\rfloor$ denotes the greatest integer which is smaller than or equal to $x$. For a given $j$, $\varphi_\omega(j)$  is a polynomial of $\omega$ with the degree of $2j-1$.  Let $\omega_{\max}$ be the largest root of $\varphi_\omega(j)$. If $ 2j-1 \gg 1$,  the oscillating behaviours of $\varphi_\omega(j)$ for $|\omega|< \omega_{\max}$ are shown in Fig. \ref{fig:plotomega1}.

\begin{figure}
\begin{center}
\includegraphics[width=0.8\textwidth]{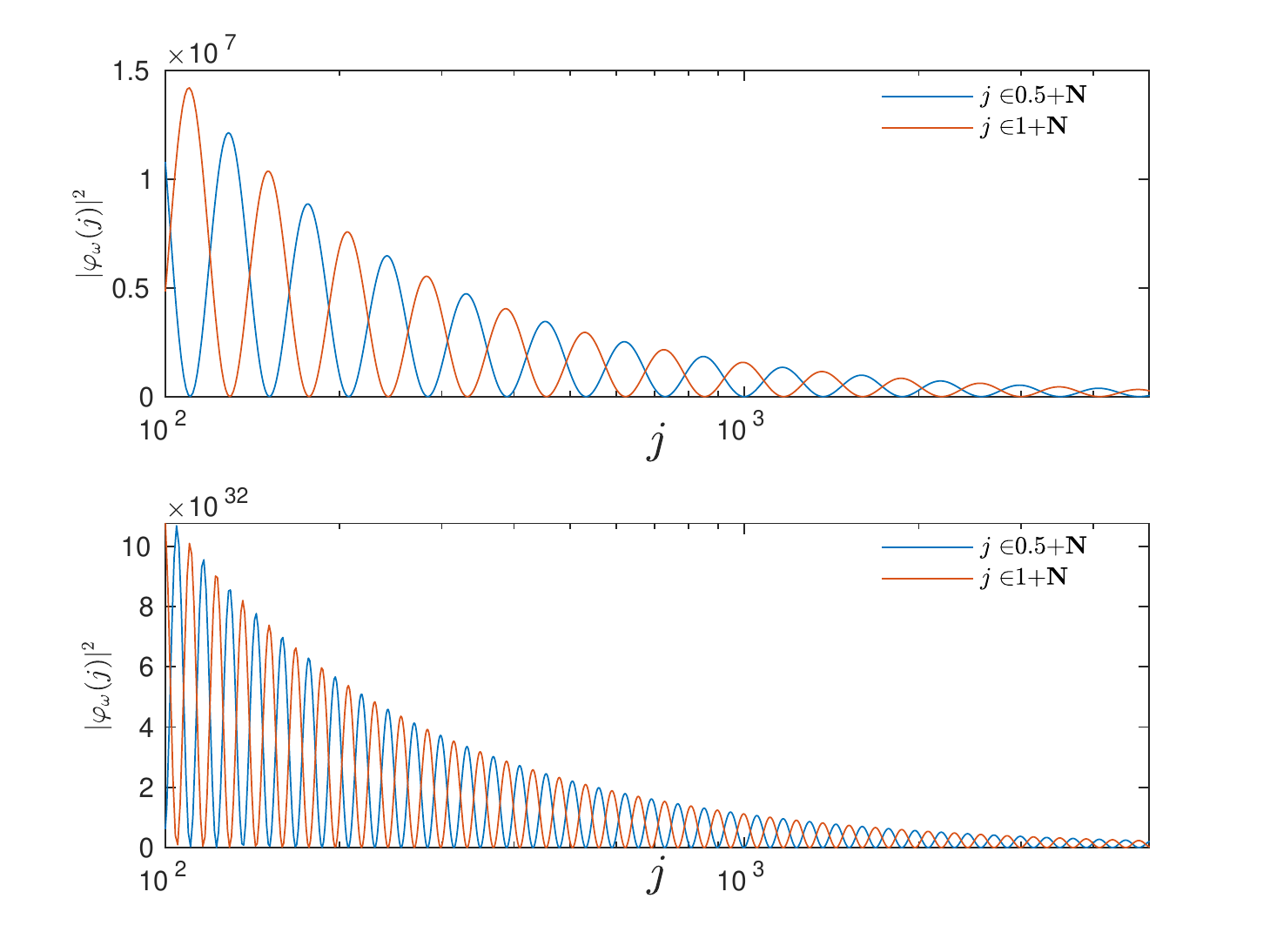}
\end{center}
\caption{The numerical results of $\varphi_\omega(j)$ in \eqref{eq:wushier} with initial data $\varphi_\omega(1/2)=1$. We let $\omega$ are $\omega=10$ for the up panel and $\omega=30$ for the other. Accrding to the results, ${\varphi}_\omega(j)$ converges to two different functions $f^\pm_\omega(j)$ as $j\to \infty$. }\label{fig:tildephi}
\end{figure}

\begin{figure}
\begin{center}
\includegraphics[width=0.8\textwidth]{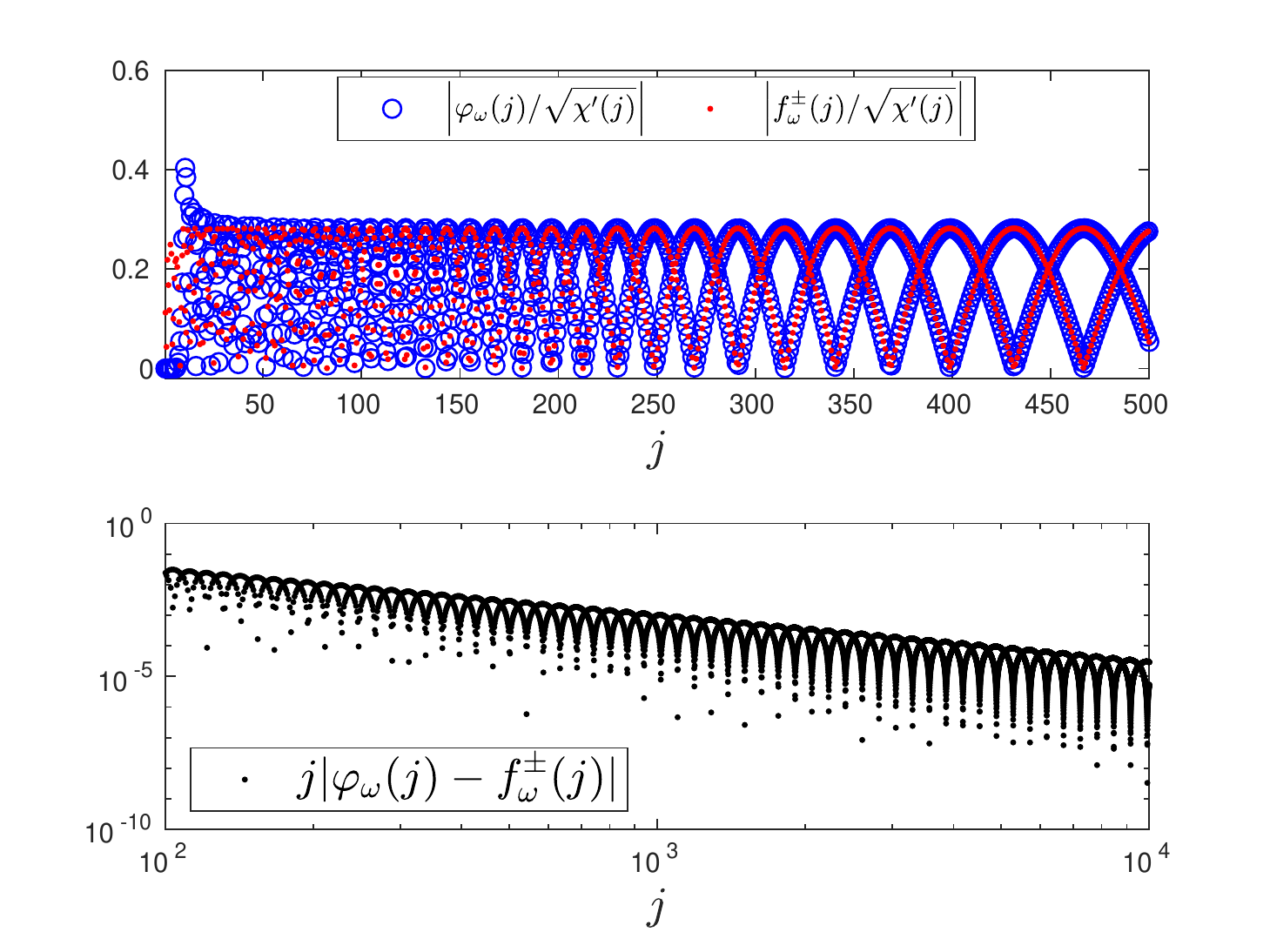}
\end{center}
\caption{A comparison between $ \varphi_\omega(j)$ and $f^\pm_\omega(j)$ for $\omega=20$.  The up panel shows the
 numerical results of $\left|\varphi_\omega(j)\right|/\sqrt{\chi'(j)}$ (blue circles) and $\left|f^\pm_\omega(j)\right|/\sqrt{\chi'(j)}$ (red
  dots) respectively. The down panel shows the numerical results of $j\left|\varphi_\omega(j)-f^\pm_\omega(j)\right|$. The evelop of $j\left|\varphi_\omega(j)-f^\pm_\omega(j)\right|$ decreases to 0 as $j$ goes to infinity, and hence $\varphi_\omega(j)$ is well estimated by $f_\omega^\pm (j)$ up to terms of order $1/j^{\Delta}$ with $\Delta>1$ for large $j$.  }\label{fig:eigenstate}
\end{figure}

Let us now focus on the asymptotic behaviours of $\varphi_\omega(j)$ for $j\gg 1/2$.
According to \eqref{eq:simplerecurrence}, in this case, $\varphi_\omega(j)$ satisfies
\begin{equation}\label{eq:wushier}
\begin{aligned}
i\omega \varphi_\omega(j)=\xi(\sqrt{(j+1/2)(j+1)}~)\varphi_\omega(j+1/2)-\xi(\sqrt{j(j+1/2)}~)\varphi_\omega(j-1/2).
\end{aligned}
\end{equation}
Because of the identity 
\begin{equation*}
\xi(\sqrt{x(x+1/2)}~)=\sqrt{\xi(x)\xi(x+1/2)}+O(1/x^3),
\end{equation*}
one can get
\begin{equation}\label{eq:largejequation}
\begin{aligned}
i\omega \varphi_\omega(j)=&\left(\sqrt{\xi(j+1/2)\xi(j+1)}+O(1/j^3)\right)\varphi_\omega(j+1/2)\\
&-\left(\sqrt{\xi(j)\xi(j+1/2)}+O(1/j^3)\right)\varphi_\omega(j-1/2).
\end{aligned}
\end{equation}
If we consider the large $j$ case in which the higher order term can be ignored, the right hand side of \eqref{eq:largejequation} can be approximated by the differentiation 
$$\sqrt{\xi(j+1/2)}\frac{\dd}{\dd j}\sqrt{\xi(j+1/2)}\varphi_\omega(j).$$
One would therefore expect the solution to the differential equation to be the asymptotics of $\varphi_\omega(j)$ for large $j$. However, the numerical analysis in Fig. \ref{fig:tildephi} tells us that $\varphi_\omega(j)$ does not converge to a single differentiable function.
It converges to a function $f_\omega^-(j)$ for the half-integers $j$ and another function $f_\omega^+(j)$ for the integers $j$ as $j$ does to infinity.  According to \eqref{eq:largejequation},  $f_\omega^\pm(j)$ satisfies 
\begin{equation}\label{eq:appromax}
\begin{aligned}
i\omega f_\omega^\pm(j)&=\sqrt{\xi(j+1/2)}\frac{\dd}{\dd j}\sqrt{\xi(j+1/2)} f_\omega^\mp(j).
\end{aligned}
\end{equation}
Solving \eqref{eq:appromax}, we get
\begin{equation}\label{eq:estimate_e}
f_\omega^\pm(j)=\sqrt{\chi'(j)}\left( A e^{i\omega \chi(j)}\mp \bar{A} e^{-i\omega \chi(j)} \right)
\end{equation}
where $\chi(x):=\ln\left(2 \sqrt{4 x^2+4 x+1}+\sqrt{16 x^2+16 x+5}\right)$ satisfies $\chi'(x)=1/\xi(x+1/2)$.  
Note that the coefficients in \eqref{eq:estimate_e} are set to be consistent to that the initial data $\varphi_\omega(1/2)$ is chosen to be real. Substituting $f^\pm_\omega(j)$ into the right hand side of \eqref{eq:largejequation}, one obtains 
\begin{equation}
\begin{aligned}
{\rm RHS}=i\omega f_\omega^\mp(j)+\frac{\sqrt{\xi(j+1/2)}}{24}g_\pm^{(3)}(c)+o((\frac{1}{j})^3)
\end{aligned}
\end{equation}
where $g_\pm(x):=\sqrt{\xi(x+1/2)}f_\omega^\pm(x)$  and $c$ is some point in  $(j-1/2,j+1/2)$. Because of $\sqrt{\xi(j+1/2)}~g^{(3)}_{\pm}(j)=O(\frac{1}{j^{5/2}})$, $f_\omega^\pm (j)$ is a asymptotic solution to \eqref{eq:wushier} up to the order $O(1/j^{5/2})$. Hence,
$\varphi_\omega(j)$ is well estimated by $f_\omega^\pm (j)$ up to terms of order $1/j^{\Delta}$ with $\Delta>1$ for large $j$. This estimation which matches with the numerical calculation. An example is shown in Fig. \ref{fig:eigenstate}.

According to \eqref{eq:estimate_e}, the function $\varphi_\omega(j)$ is bounded as $j$ goes to infinity. However, it is not normalizable.
To show this, let us fix a large integer $j_0$. Then we have 
\begin{equation}\label{eq:normeigen}
\begin{aligned}
\sum_{j=1/2}^{\infty}|\varphi_\omega(j)|^2\geq & \sum_{j=1/2}^{j_0-1/2}\left|\varphi_\omega(j)\right|^2+\sum_{ j=j_0,j_0+1\cdots}\left|f_\omega^+(j)-\varphi_\omega(j)\right|^2+\sum_{ j=j_0+1/2,j_0+3/2\cdots }\left|f_\omega^-(j)-\varphi_\omega(j)\right|^2\\
&-\sum_{ j=j_0,j_0+1\cdots}2\left|f^+_\omega-\varphi_\omega(j)\right|\left|f_\omega^+(j)\right|-\sum_{ j=j_0+1/2,j_0+3/2\cdots }2\left|f^-_\omega-\varphi_\omega(j)\right|\left|f_\omega^-(j)\right|\\
&+\sum_{ j=j_0,j_0+1\cdots}\left|f_\omega^+(j)\right|^2+\sum_{ j=j_0+1/2,j_0+3/2\cdots }\left|f_\omega^-(j)\right|^2
\end{aligned}
\end{equation}
Because $\left|\varphi_\omega(j)-f_\pm(j)\right|$ decreases as $1/j^{\Delta}$ with $\Delta>1$, we have
\begin{equation*}
\begin{aligned}
\psi_0:=&\sum_{j=1/2}^{j_0-1/2}\left|\varphi_\omega(j)\right|^2+\sum_{ j=j_0,j_0+1\cdots}\left|f_\omega^+(j)-\varphi_\omega(j)\right|^2+\sum_{ j=j_0+1/2,j_0+3/2\cdots }\left|f_\omega^-(j)-\varphi_\omega(j)\right|^2\\
&-\sum_{ j=j_0,j_0+1\cdots}2\left|f^+_\omega-\varphi_\omega(j)\right|\left|f_\omega^+(j)\right|-\sum_{ j=j_0+1/2,j_0+3/2\cdots }2\left|f^-_\omega-\varphi_\omega(j)\right|\left|f_\omega^-(j)\right|< \infty
\end{aligned}
\end{equation*}
Therefore, we get
\begin{equation}\label{eq:fourfour}
\sum_{j=1/2}^{\infty}|\varphi_\omega(j)|^2\geq \psi_0+ \sum_{ j=j_0,j_0+1\cdots}|f_+(j)|^2+\sum_{ j=j_0+1/2,j_0+3/2\cdots }|f_-(j)|^2.
\end{equation}
The remaining terms in the right hand side of \eqref{eq:fourfour}  apart from $\psi_0$ are infinity because of
$$
\int_{j_0}^\infty |f_\pm(x)|^2 \dd x=\infty.  
$$
In conclusion,  $|\omega\rangle$ is not normalizable since $\langle \omega|\omega\rangle=\infty$. Thus it is not well defined within the Hilbert space $\mathcal{H}_\gamma$. However, the function $\varphi_\omega(j)$ can be cut-off properly to get a sequence of states $|\omega\rangle_n\in\mathcal{H}_\gamma$, such that ${}_n\langle\omega|\omega\rangle_n=1$ and $\lim_{n\to\infty}(\hat{H}_I-\omega)|\omega\rangle_n=0$. Such a sequence is known as the Weyl sequence. Given an $\omega$, the existence of the Weyl sequence converging to $|\omega\rangle$ ensures the fact that $\omega\in\sigma(\hat{H}_I)$ \cite{reed1975}.
The rigorous mathematics of this procedure is shown in the appendix \ref{app:weyl}, where we show that for each $\omega\in\mathbb{R}$ there exists a Weyl sequence.  It is remarkable that, for $\omega=0$ there exists not only this approximate eigenstate, i.e. the Weyl sequence, but also an exact eigenstate $|j=0\rangle$. In the following contexts, we will denote $|j=0\rangle=:|\oo\rangle$ and $|\omega=0\rangle=:|0\rangle$ to distinguish these two states. It is easy to get $\langle 0|\oo\rangle=0$. We finally conclude the following theorem, 
\begin{theorem}
The spectrum of $\hat{H}_I$ is $\mathbb{R}$, i.e.  $\sigma(\hat{H}_I)=\mathbb{R}$, and $0\in\sigma(\hat{H}_I)$ is the point spectrum. 
\end{theorem}   

Since $\hat{H}_I$ is diagonalized in the basis $\dualvec{\omega}$\footnote{ More precisely, $|\omega\rangle$ is in the dual space $\f^*\supset\mathcal{H}_\gamma\supset\f$ rather than $\mathcal{H}_\gamma$. }, 
the orthogonality relation is given by
\begin{equation}
\begin{aligned}
\langle \omega'\dualvec{\omega}=\delta(\omega',\omega)
\end{aligned}
\end{equation}
By using the basis of $\dualvec{\omega}$, a state in $\mathcal{H}_\gamma/\{|\oo\rangle\}$ can always be decomposed as 
\begin{equation}\label{eq:initialstate}
\dualvec{\Psi}=\int_{\mathbb{R}}\dd\omega \Psi(\omega)\dualvec{\omega}.
\end{equation}
 Then the inner product between two states reads 
\begin{equation}
\langle\Psi_1\dualvec{\Psi_2}=\int_{\mathbb{R}}\dd\omega \overline{\Psi_1(\omega)}\Psi_2(\omega).
\end{equation}

\begin{figure}
\begin{center}
\includegraphics[width=0.5\textwidth]{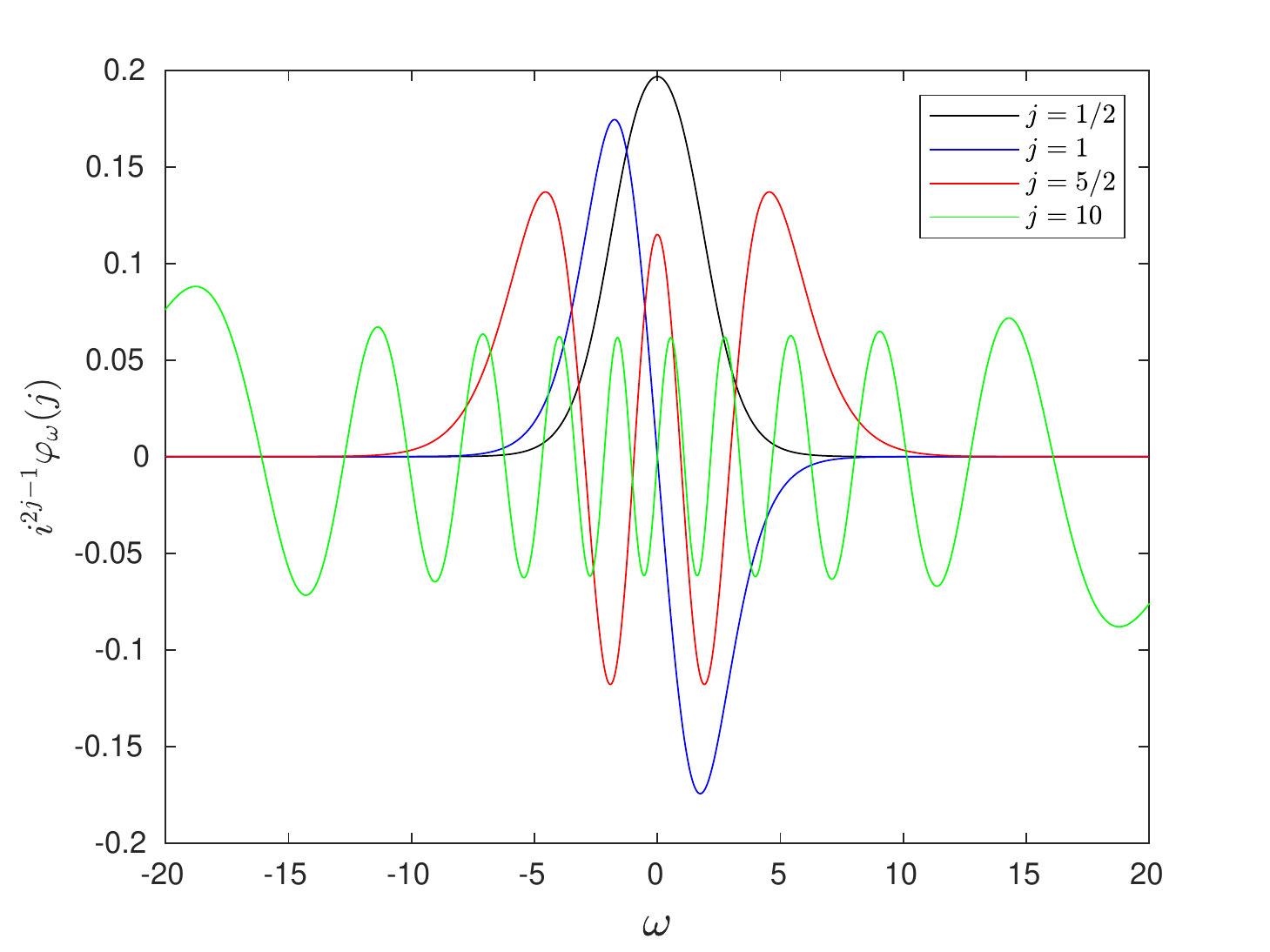}
\end{center}
\caption{Plot of normalized $\varphi_\omega(j)$ as a function of $\omega$. As shown by the black line, the initial data $\varphi_\omega(1/2)$ is a Gaussian-like function of $\mathcal{N}e^{-\mathfrak{J}(\omega)}$. }\label{fig:ini}
\end{figure}

We now discuss the issue of the normalization of $\varphi_\omega(j)$. If $\varphi_\omega(j)$ is normalized, it satisfies both
\begin{equation}\label{eq:normfirst}
\int_\mathbb{R}\overline{\varphi_\omega(j)}\varphi_\omega(j')\dd\omega=\delta_{j,j'}
\end{equation}
and
\begin{equation}\label{eq:normsecond}
\sum_{j}\overline{\varphi_\omega(j)}\varphi_{\omega'}(j)=\delta(\omega,\omega').
\end{equation}
However, the $\varphi_\omega(j)$ in \eqref{eq:varphiformula}, obtained by choosing the initial condition $\varphi_\omega(1/2)=1$, satisfies neither  \eqref{eq:normfirst} nor \eqref{eq:normsecond}.  To normalize the present $\varphi_\omega(j)$, we should re-set the initial data $\varphi_\omega(1/2)$ by finding a weight function $w(\omega)$ such that
\begin{equation}
\int_\mathbb{R}\overline{\varphi_\omega(j)}\varphi_\omega(j')w(\omega)\dd\omega =\delta_{j,j'}
\end{equation}
Then the initial data can be set as $\varphi_\omega(1/2)=\sqrt{w(\omega)}$. Though this procedure  is mathematically rigorous,it is not practicable. A similar puzzle also appeared in loop quantum cosmology. We can use the technique was introduced in \cite{kaminski2010cosmic} to deal with this difficulty. We refer to appendix \ref{app:normalization} for details of the calculation, according to which the initial data should be chosen such that the factor $A$ appearing in the asymptotic formula \eqref{eq:estimate_e} of $\varphi_\omega(j)$ satisfies $|A|=\frac{1}{2\sqrt{\pi}}$. This can be done numerically as shown in Fig. \ref{fig:ini}, according to which the initial data is chosen as some Gaussian-like function, denoted as $\varphi_\omega(1/2)=\mathcal{N}e^{-\mathfrak{J}(\omega)}$.

From now on, $\varphi_\omega(j)$ is referred refers to the normalized version. In other words, $\varphi_\omega(j)$ in \eqref{eq:varphiformula} is re-defined as
\begin{equation}\label{eq:revarphiformula}
\varphi_\omega(j)=\mathcal{N}e^{-\mathfrak{J}(\omega)}\sum_{\mu=0}^{\lfloor j-\frac{1}{2}\rfloor}\zeta_{j,\mu}\omega^{2j-1-2\mu}, 
\end{equation}
and the asymptotic behaviour of \eqref{eq:estimate_e} is rewritten as  
\begin{equation}\label{eq:reestimate_e}
f_\omega^\pm(j)=\frac{\sqrt{\chi'(j)}}{2\sqrt{\pi}}\left( e^{i\omega \chi(j)+i\theta(\omega)}\mp  e^{-i\omega \chi(j)-i\theta(\omega) }\right)
\end{equation}
where $\theta(\omega)$ is the phase of $A$ in \eqref{eq:estimate_e}, namely $A=|A|e^{i\theta(\omega)}$.

\section{Dynamics driven by $\hev$}\label{se:dynamic}
 In this section we will discuss the dynamic of the present model given merely by $\sqrt{\hev}$. 
Recall that in the expression $\hev=\sqrt{\hh}=\sqrt{\hat{H}_I^2+\hat{H}_R^2}$, $\hat{H}_I$ is unbounded but $\hat{H}_R$ is bounded with $||\hat{H}_R||\leq 3/2$. Thus we will regard $\hat{H}_R^2$ as a bounded perturbation on $\hat{H}_I^2$ and consider the dynamics driven by the zero-order of $\hev$, i.e. by $|\hat{H}_I|$. As before, the state $|\oo\rangle$ should be specially treated. This state is preserved by the evolution operator $e^{it\sqrt{\hev}}$ in \eqref{eq:evoop}. Trivial results can also be obtained if the actions of observables like volume, area and energy $\pi^2=\hat{H}$ on this state are considered. Therefore, without loss of generality, we will restrict ourself on the subspace $\overline{\mathcal{H}_\gamma/\{|\oo\rangle\}}$.

The evolution of an initial state $\dualvec{\Psi}$ of \eqref{eq:initialstate} reads  
 \begin{equation}
 \dualvec{\Psi_t}=\int_{\mathbb{R}}\dd\omega \Psi(\omega)e^{-it \sqrt{|\omega|}}\dualvec{\omega}.
 \end{equation}
For the wave function of a coherent state
\begin{equation}
 \Psi(\omega)=e^{-\frac{(\omega-\omega_0)^2}{2\sigma^2}+i\phi\omega}
 \end{equation} 
we have
 \begin{equation}
 \begin{aligned}
 \dualvec{\Psi_t}=&\int_{\mathbb{R}}\dd \omega e^{-\frac{(\omega-\omega_0)^2}{2\sigma^2}+i\phi\omega-it\sqrt{|\omega|} }\dualvec{\omega}.
 \end{aligned}
 \end{equation}
 By using the the spin network representation, we have
 \begin{equation}\label{eq:apppsiunderj}
 \langle j\dualvec{\Psi_t}= \int_\mathbb{R} \dd\omega e^{-\frac{(\omega-\omega_0)^2}{2\sigma^2}+i\phi\omega-it\sqrt{|\omega|} }\varphi_{\omega}(j).
 \end{equation}

 \subsection{Calculation of a coherent state in spin network representation}
 \subsubsection{The case of $j\gg 1$}
 Recall that $\varphi_\omega(j)$ can be well approximated by 
 \begin{equation}\label{eq:fiveseven}
 \varphi_\omega(j)\cong \frac{\sqrt{\chi'(j)}}{2\sqrt{\pi}}( e^{i\omega\chi(j)+i\theta(\omega)}+(-1)^{2j+1}e^{-i\omega\chi(j)-i\theta(\omega)}).
 \end{equation} 
One can substitute it into \eqref{eq:apppsiunderj} to obtain 
\begin{equation}\label{eq:psijjk}
\langle j\dualvec{\Psi_t}=:\psi_t(j)\cong \frac{1}{2\sqrt{\pi}} e^{i\phi\omega_0-it \sqrt{\omega_0}}\sqrt{\chi'(j)}I_{j} 
\end{equation}
where $I_j$ is defined by
\begin{equation}\label{eq:basicint}
\begin{aligned}
I_j=&\int_{-\infty}^\infty \dd\omega e^{-\frac{(\omega-\omega_0)^2}{2\sigma^2}+i(\phi-\frac{t}{2\sqrt{\omega_0}}) ( \omega-\omega_0)}\left( e^{i\omega\chi(j)+i\theta(\omega)}+(-1)^{2j+1} e^{-i\omega\chi(j)-i\theta(\omega)}\right)\\
\cong&\sum_{s=\pm 1}s^{2j+1}\sqrt{2\pi\sigma^2}e^{ -\frac{\sigma^2\left(\phi-\frac{t}{2\sqrt{2\omega_0}}+s\chi(j) +s\theta'(\omega_0)\right)^2}{2}+is\left(\theta(\omega_0)+\chi(j)\omega_0\right)}.
\end{aligned}
\end{equation}
Note that in the second step we expanded $\theta$ around $\omega_0$. 
It is easy to obtain
\begin{equation}\label{eq:absI}
\begin{aligned}
|I_j|^2=&2\pi\sigma^2\left(e^{-\sigma^2\left(\phi-\frac{t}{2\sqrt{\omega_0}}+\theta'(\omega_0)+\chi(j)\right)^2}+e^{-\sigma^2\left(\phi-\frac{t}{2\sqrt{\omega_0}}-\theta'(\omega_0)-\chi(j)\right)^2}\right)\\
&+4\pi\sigma^2 e^{-\sigma^2(\phi-\frac{t}{2\sqrt{\omega_0}})^2-\sigma^2 (\theta'(\omega_0)+\chi(j) )^2}\cos\left(2(\theta(\omega_0)+\chi(j)\omega_0)\right).
\end{aligned}
\end{equation}
 For $j\gg 1$ such that $\chi(j)\gg 1$, we have
\begin{equation}\label{eq:sixone}
 e^{-\sigma^2(\phi-\frac{t}{2\sqrt{\omega_0}})^2-\sigma^2 (\theta'(\omega_0)+\chi(j) )^2}\ll 1.
 \end{equation} 
 For any given $j$ such that condition \eqref{eq:sixone} is satisfied, there are two possible values of $t$ where $|I_j|^2$ peaks, and can be well approximated by
 \begin{equation}\label{eq:largejIj}
 |I_j|^2\cong 2\pi\sigma^2\left(e^{-\sigma^2\left(\phi-\frac{t}{2\sqrt{2\omega_0}}+\theta'(\omega_0)+\chi(j)\right)^2}+e^{-\sigma^2\left(\phi-\frac{t}{2\sqrt{2\omega_0}}-\theta'(\omega_0)-\chi(j)\right)^2}\right).
 \end{equation}

\subsubsection{The case of $j\sim 1$}
According to \eqref{eq:revarphiformula}, $\varphi_\omega(j)$ is a polynomial of $\omega$ multiplied by a Gaussian-like function $e^{-\mathfrak{J}(\omega)}$.  
Together with \eqref{eq:apppsiunderj}, one can see that the evolution of the wave function is determined by the integral
\begin{equation*}
\begin{aligned}
I_j:=&\int_\mathbb{R}\dd\omega e^{-\frac{(\omega-\omega_0)^2}{2\sigma^2}+i\left(\phi-\frac{t}{2\sqrt{2\omega_0}}\right)(\omega-\omega_0)}\varphi_\omega(j)\\
=&\mathcal{N}\sum_{\mu=0}^{\lfloor j-\frac{1}{2}\rfloor}\zeta_{j,\mu}\int_\mathbb{R}\dd\omega e^{-\frac{(\omega-\omega_0)^2}{2\sigma^2}-\mathfrak{J}(\omega)+i\left(\phi-\frac{t}{2\sqrt{2\omega_0}}\right)(\omega-\omega_0)}\omega^{2j-1-2\mu}.
\end{aligned}
\end{equation*}
We expand $\mathfrak{J}$ around $\omega_0$ to perform the integration as in the appendix \ref{app:saddle} and obtain
\begin{equation}
\begin{aligned}
I_j\cong &\sqrt{2\pi}\sigma e^{\frac{\sigma^2}{2}\left(i\phi-\frac{i t}{2\sqrt{\omega_0}}-\mathfrak{J}'(\omega_0)\right)^2}\Big(\varphi_{\mathring{\omega}}(j)+\Delta\Big)
\end{aligned}
\end{equation}
where 
\begin{equation}
\begin{aligned}
\mathring{\omega}&=\omega_0+\sigma^2\left(i\phi-\frac{i t}{2\sqrt{\omega_0}}-\mathfrak{J}'(\omega_0)\right),\\
\Delta&=e^{-\mathfrak{J}(\omega_0)}\mathcal{N}\sum_{\mu=0}^{\lfloor j-\frac{1}{2}\rfloor}\zeta_{j,\mu}\sum_{m=1}^{\lfloor j-\mu-\frac{1}{2} \rfloor}\frac{(2j-2\mu-1)!\sigma^{2m}\mathring{\omega}^{2j-2\mu-1-2m}}{m!(2j-2\mu-1-2m)!2^m}.
\end{aligned}
\end{equation}
and we used the formula
\begin{equation*}
\int_\mathbb{R}e^{-\frac{(x-x_0)^2}{2\sigma^2}}x^n\dd x=\sqrt{2\pi}\sigma\sum_{m=0}^{\lfloor \frac{n}{2}\rfloor}\frac{n!\sigma^{2m} (x_0)^{n-2m}}{m!(n-2m)!2^m}.
\end{equation*}
By using the fact that $\omega_0\gg 1$, we have
\begin{equation}
\begin{aligned}
\left|\frac{\Delta}{\varphi_{\mathring{\omega}}(j)}\right|&=\sum_{m=1}^{\lfloor j-\frac{1}{2}\rfloor}\frac{(2j-[[k]])!\sigma^{2m}}{m!(2j-1-2m)!2^m|\mathring{\omega}|^{2m}}+O(1/|\mathring{\omega}|^2)\\
&\leq \sum_{m=1}^{\lfloor j-\frac{1}{2}\rfloor} \frac{(2j-1)^{2m}\sigma^{2m}}{2^m|\mathring{\omega}|^{2m}}+O(1/|\mathring{\omega}|^2).
\end{aligned}
\end{equation}
For $\frac{(2j-1)\sigma }{\sqrt{2}|\mathring{\omega}|}\ll 1$, we have $\Delta\ll | \varphi_{\mathring{\omega}}(j) |$, and hence
 \begin{equation}
 I_j\cong \sqrt{2\pi}\sigma e^{\frac{\sigma^2}{2}\left(-(\phi-\frac{t}{2\sqrt{2\omega_0}})^2+\mathfrak{J}'(\omega_0)^2-2i\mathfrak{J}'(\omega_0)(\phi-\frac{t}{2\sqrt{2\omega_0}})\right)}\varphi_{\mathring{\omega}}(j). 
 \end{equation}
Because $\omega_0$ is equal to $\mathring{\omega}$ up to a term of order $\sigma^2$, we get
\begin{equation}\label{eq:smalljIJ}
|I_j|\cong \sqrt{2\pi}\sigma e^{\frac{\sigma^2}{2}\mathfrak{J}'(\omega_0)^2} e^{-\frac{\sigma^2}{2}(\phi-\frac{t}{2\sqrt{2\omega_0}})^2}|\varphi_{\omega_0}(j)|.
\end{equation}

\subsection{Physical analysis of the results}\label{se:physicalanalysis}

\begin{figure}
\begin{center}
\includegraphics[width=0.8\textwidth]{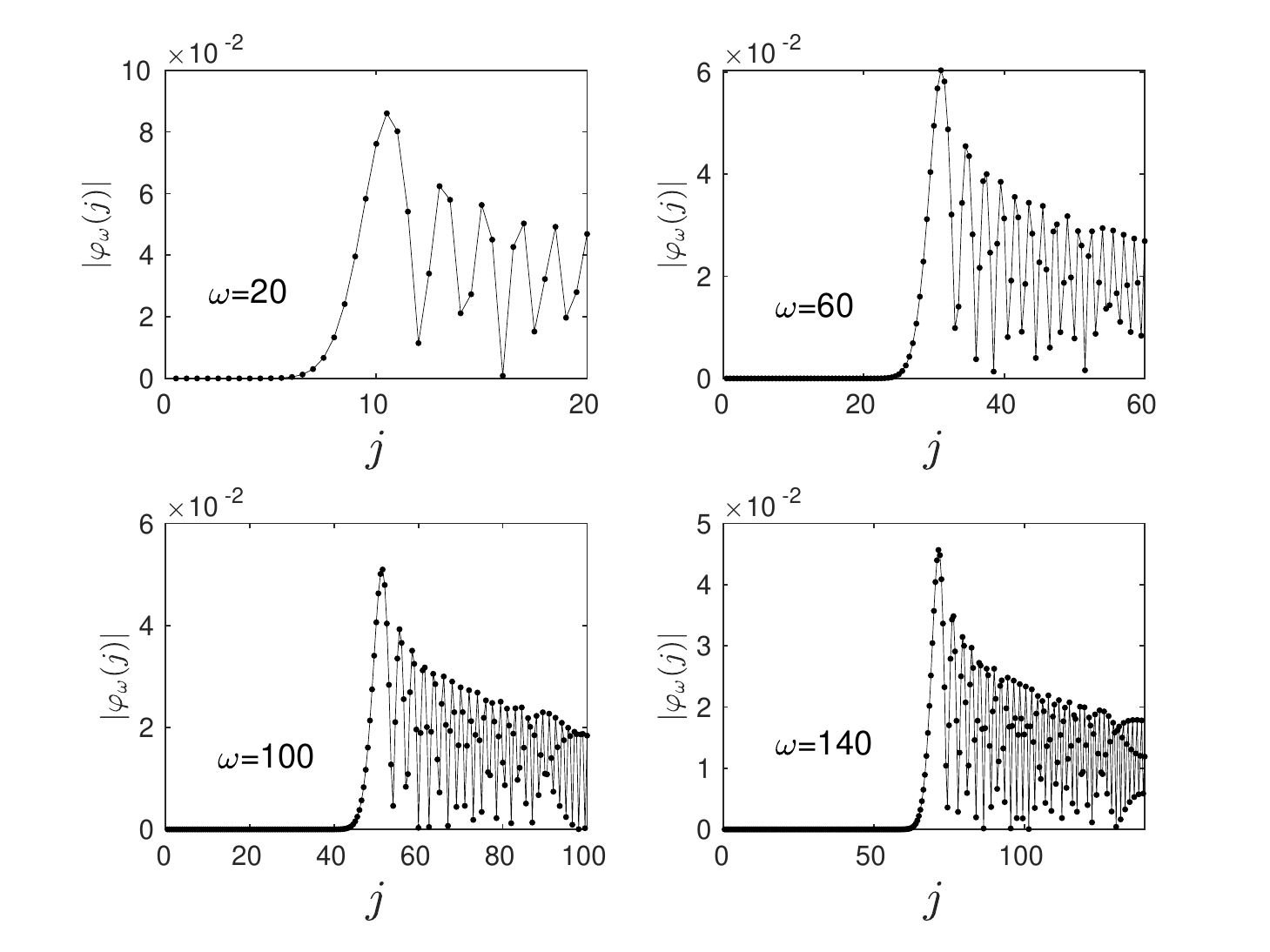}
\end{center}
\caption{Plot of eigenfunction $\varphi_\omega(j)$. For each $\omega$, there is always a barrier at some fixed spin $j$ where the bounce happens. }\label{fig:wall}
\end{figure}

Let us now analyse the results of \eqref{eq:largejIj} and \eqref{eq:smalljIJ}. 
In the case of $|\phi-\frac{t}{2\sqrt{\omega_0}}|\gg 1$, \eqref{eq:largejIj} implies that $|I_j|^2$ takes an extremum at $j_{0}$ in the large $j$ region. For $j\sim 1$, \eqref{eq:smalljIJ} implies $|I_j|^2\propto|\varphi_\omega(j)|^2$, and numerical analysis shows that $|\varphi_\omega(j)|^2\ll 1$ (examples are given in figure \ref{fig:wall}). Therefore the extremum of $|I_j|$ is the global maximum. This means that the state $\psi_t(j)$ is concentrated on the large $j$ region for $|\phi-\frac{t}{2\sqrt{2\omega_0}}|\gg 1$. \eqref{eq:largejIj} implies that we could think $\psi_t(j)$ as a wave packet resembling the evolution of a classical gravity system. For a given spin $j\gg 1$, the right hand side of \eqref{eq:largejIj} consists of two Gaussian functions of $t$. Therefore, the wave packet will reach a fixed spin $j$ twice during its evolution at two different moments $t_0^\pm$ given by\footnote{In order to obtain this picture, it is necessary to require the spin $j$ to satisfy $|t_0^+-t_0^-|>2\sqrt{\omega_0}/\sigma$, so that the peaks of the two Gaussian functions are separated.}
 \begin{equation}\label{eq:t0pm}
 t_0^\pm=2\sqrt{2\omega_0}\left(\phi\pm\theta'(\omega_0)\pm\chi(j)~\right).
 \end{equation}
This behaviour already implies to us a picture of bounce evolution roughly.

\begin{figure}
\centering
\includegraphics[width=0.6\textwidth]{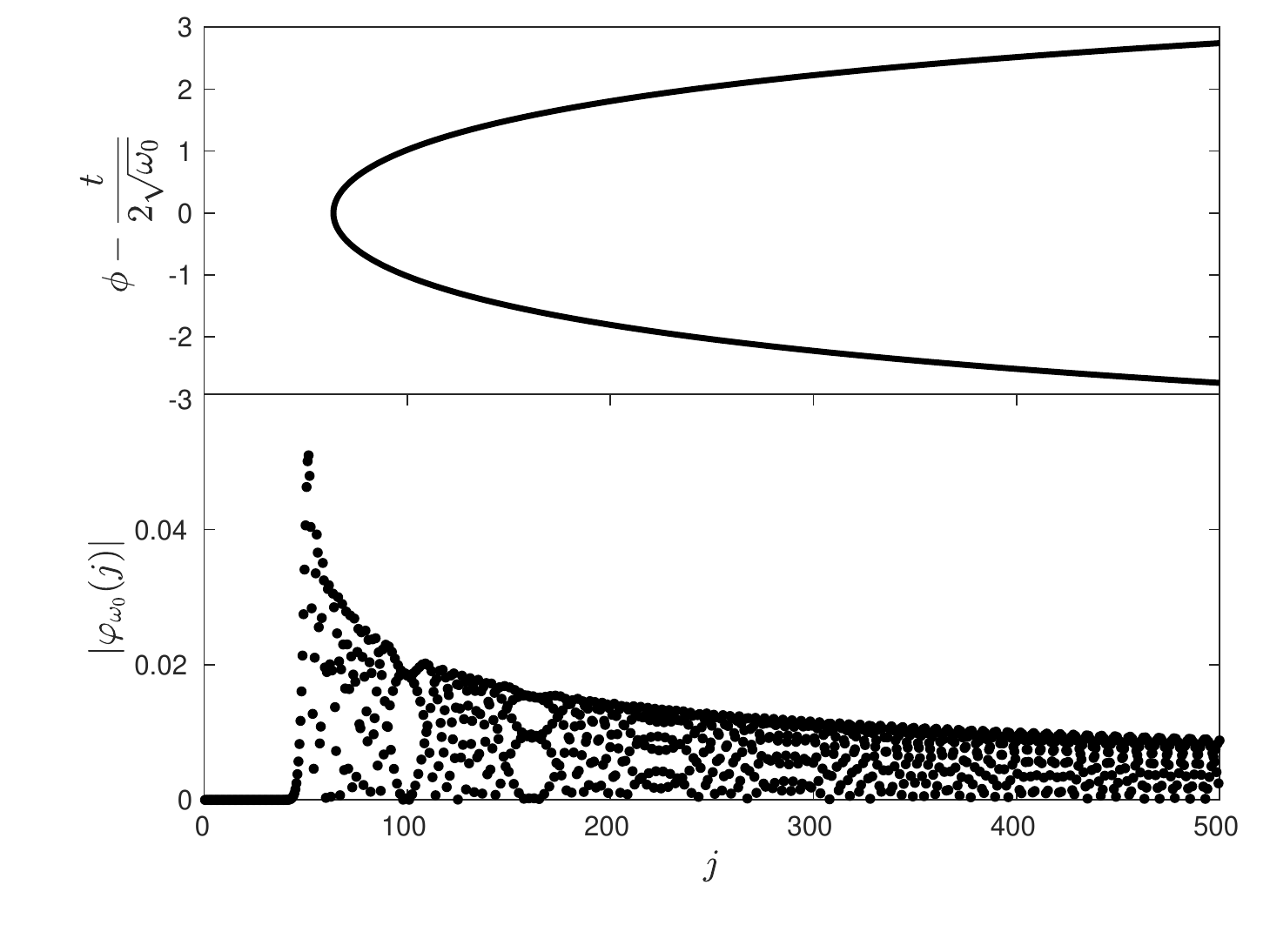}
\caption{Plot of evolutions of the expectation value $\sum\limits_jj|\psi_t(j)|^2$ with the state $\psi_t(j)$ given in \eqref{eq:apppsiunderj} (up panel) and wave function $|\varphi_{\omega_0}(j)|$(down panel). The parameters are chosen as: $\sigma=1$, $\omega_0=100$.  The bounce happens close to the `barrier'.}\label{fig:expj}
\end{figure}

\begin{figure}
\centering
\includegraphics[width=0.6\textwidth]{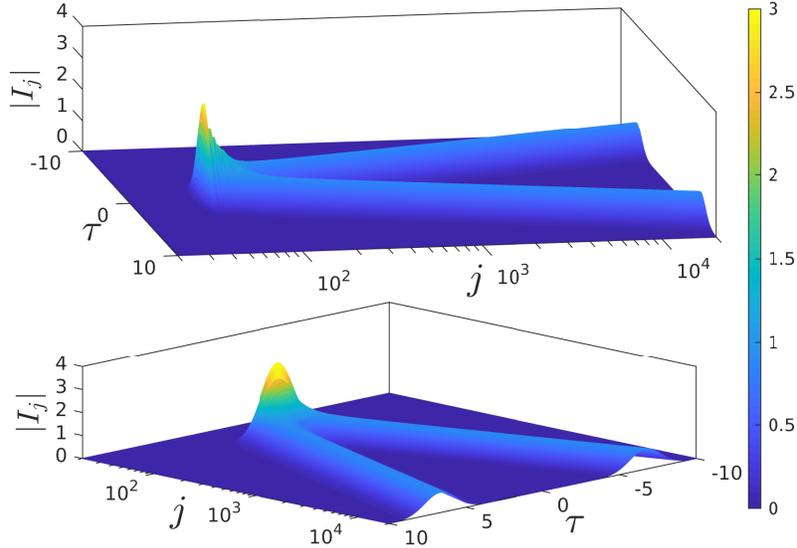}
\caption{Numerical results of $I_j:=\sqrt{\chi'(j)}\psi_t(j)$,  visualized from different viewing angles, where we choose $\sigma=1$, $\omega_0=100$. The $\tau$-axis corresponds to the phase $(\phi-\frac{t}{2\sqrt{\omega_0}})$ in \eqref{eq:apppsiunderj}.}\label{fig:IJ}
\end{figure}

In the case of $|\phi-\frac{t}{2\sqrt{\omega_0}}|\sim 1$, \eqref{eq:largejIj} implies that $|I_j|^2$ decays exponentially in the large $j$ region. Hence, physics of the system is concentrated in the small $j$ region, that is the quantum region. Equation \eqref{eq:smalljIJ} implies $|I_j|^2\propto|\varphi_{\omega_0}(j)|^2$ when $j$ is relatively small. Hence the dynamical behaviour of this system, given by $I_j$, is intrinsic in the eigenfunction itself. Numerical calculations reveal the properties of $\varphi_\omega(j)$ as follows. As shown in figure \ref{fig:wall}, a jump of $|\varphi_\omega(j)|$ will always occur when $j$ reaches certain value, where $|\varphi_\omega(j)|$ as well as $|I_j|$ increases sharply from some value much smaller than $1$ in the small $j$ region. Because of this jump of $|\varphi_\omega(j)|$, a barrier is formed at some $j_b^{(\omega)}$ by the eigenfunction of the Hamiltonian itself, which blocks physical states from reaching a smaller $j$ such as $j_{\rm min}=1/2$. This barrier is where the bounce of the evolution occurs. Some numerical results are shown in figures \ref{fig:expj} and \ref{fig:IJ}. It is shown in figure \ref{fig:expj} that a bounce happens close to the `barrier'. In figure \ref{fig:IJ}, a numerical result of $I_j:=\sqrt{\chi'(j)}\psi_t(j)$ is shown ao that the evolution of the wave packet is visualized.
 

We now discuss the effect of the factor $\sqrt{\chi'(j)}$ in \eqref{eq:psijjk}. By definition, one has $\chi'(j)\sim 1/j$ which will depress the value of $\psi_t(j)$ for $j\gg 1$. Then even though $|I_j|$ has a global maximum at some $j_0\gg 1$, this may not be the case for $|\psi_v(j)|$ due to the factor $\sqrt{\chi'(j)}$. However, we will see below that it is $I_j$ rather than $\psi_t(j)$ which determines the evolution of the expectation value of a geometrical observable under the coherent state.
Consider an operator $\mathcal{F}$ which can be diagonalized under the spin network basis as
\begin{equation*}
\mathcal{F}\dualvec{j,k}=f(\chi(j))\dualvec{j}. 
\end{equation*}
The expectation value $\langle\mathcal{F}\rangle$ can be calculated as
\begin{equation}\label{eq:summationf}
\langle \mathcal{F}\rangle=\mathcal{N}\sum_{j}f(\chi(j))\left|\psi_t(j)\right|^2
\end{equation}
where $\mathcal{N}$ is the normalization factor. As discussed above, in the case of $|\phi-\frac{t}{2\sqrt{\omega_0}}|\sigma\gg 1$, the summation \eqref{eq:summationf} is dominated by the values in $\psi_t(j)$ of large $j$ region. Thus the approximation \eqref{eq:psijjk} leads to
\begin{equation}
\begin{aligned}
\langle\mathcal{F}\rangle\cong &\mathcal{N}\sum_{\text{large }j} f(\chi(j))\chi'(j)|I_{j}|^2.
\end{aligned}
\end{equation}
Since $I_j$ takes the form $I_j=G(\chi(j))$ with $G(x)$ a Schwartz function according to \eqref{eq:largejIj} and $\chi(j)\sim \ln(j)$, we can approximate the summation by integral as
\begin{equation}\label{eq:expf}
\langle \mathcal{F}\rangle\cong\mathcal{N}\int \dd xf(x) |G(x)|^2.
\end{equation}
Hence, the expectation value $\langle \mathcal{F}\rangle$ is determined by $G(x)$, namely $I_j$, rather than $\varphi_\omega(j)$. For instance if we choose $\mathcal{F}$ as $\mathcal{F}\dualvec{j}:=j\dualvec{j}$, i.e. $f(\chi(j))=j$, the evolution of the expectation value of $\mathcal{F}$, which is $\sum\limits_jj|\psi_t(j)|^2$, can be calculated numerically, as in figure \ref{fig:expj}.

\section{Summary and Remark}\label{se:conclution}
To summarize, in previous sections, in order to understand the dynamics of LQG, we apply a symmetrized version of the graph-preserving Hamiltonian operator \eqref{theoperator} in LQG on a simple graph $\gamma$ containing one loop based at a single vertex. In the corresponding sub-Hilbert-space $\mathcal{H}_\gamma$,  the action of the Hamiltonian operator is calculated by choosing a proper spin network basis. It is proven that the Hamiltonian operator restricted on $\mathcal{H}_\gamma$ is self-adjoint and its spectrum is the entire real line. Based on these results, the dynamics is analysed in details. A remarkable result as discussed in section \ref{se:physicalanalysis} is that a picture of bouncing evolution can be obtained.

As shown in  figure \ref{fig:expj} and figure \ref{fig:IJ}, the picture of bouncing evolution is depicted by several properties of the time dependent wave function $\psi_t(j)$ of \eqref{eq:apppsiunderj}.  First, for $|t|\gg 1$, $\sqrt{\chi'(j)}|\psi_t(j)|$ will be kept as a Gaussian-like function of $\chi(j)$ during evolution. Thus this wave packet may represent a state which has the dynamics closely resembling some classical gravity system. Then, if we fix spin $j\gg 1$ and consider $\psi_t(j)$ as a function of time $t$, $|\psi_t(j)|^2$ (or $|I_j|^2$ equivalently) distributes on variable $t$ as a summation of two Gaussian functions as in \eqref{eq:largejIj}. Therefore, the wave packet reaches the given spin $j$ twice during its evolution at two different moments $t_0^\pm$ in \eqref{eq:t0pm}. Finally, in the region around $j_{\rm min}=1/2$, we always have $|\psi_t(j)|^2\ll 1$. Therefore the wave packet is bounced from some spin $j>j_{\rm min}$.  

As shown in section \ref{se:dynamic}, the above properties of $\psi_t(j)$ come from the properties of the eigen-functions $\varphi_\omega(j)$ of the Hamiltonian operator. First, in the large $j$ region, the functions $\varphi_\omega(j)$ are given by superposition of an incoming wave $e^{i \omega\chi(j)}$ and an outgoing wave $e^{-i\omega\chi(j)}$ in \eqref{eq:fiveseven}. This property leads to the fact that $\psi_t(j)$ is given by a summation of two Gaussian function as \eqref{eq:largejIj}. Second, in the small $j$ region, the eigen-functions increase sharply to form a barrier which blocks $\psi_t(j)$ and geometrical observables from reaching the minimum spin $j_{\rm min}$ as shown in figures \ref{fig:wall} and \ref{fig:expj}. The bounce happens where the barrier appears. Hence the barrier is the key element to cause the bouncing evolution. Our numerical investigation shows that, this kind of barrier can always appear if the eigen-equation of the Hamiltonian operator under spin network basis, like \eqref{eq:recurrence}, is a difference equation and the coefficients in the combination of the right hand side takes the form $j^n$ with $n\neq 0$ in general. Since difference equation is a result of the discreteness of the spacetime geometry,  in this sense the singularity resolution of model is related to the discreteness of quantum spacetime geometry.

It should be noted that in the present work the dynamics of a graph-preserving physical Hamiltonian in LQG is studied as an example on the states based on one loop at a point. This model can be regarded as a homogeneous sector  of full LQG theory where the entire quantum spacetime is coarse-grained as a single vertex. This picture mimics a cosmological sector of GR where all of the degrees of freedom are reduced to a point. Surprisingly, this solution, arising completely from full LQG theory, shares similar dynamical behaviours as LQC, although this result is obtained through a completely different approach.

Several remarks on the limitation of the present work are listed below. First, we considered the dynamic driven by $\sqrt{\hh}$, or essentially by the particular symmetric version $\hat{H}_I$ defined by the second equation in \eqref{eq:HRI}. If one considered the other symmetric version $\hat{H}_R$ defined by the first equation in \eqref{eq:HRI}, there would not appear a barrier in its eigen-functions. Then a picture of bouncing evolution could not be expected. Second, taking into account of the fact that the Lorentzian part of the Hamiltonian operator is still a rather open issue, we only considered dynamics of the Euclidean part of the Hamiltonian operator. It is unclear at the moment whether the Lorentzian part would play an important role for the bouncing evolution. All of these open issues are left for future study. 

\section{Acknowledgement}
This work was
supported by the Natural Science Foundation of China (NSFC), grand No. 11875006  and the  Polish Narodowe Centrum Nauki, Grant No. 2018/29/B/ST2/01250.

\appendix
\section{Action of $\hek$ }\label{app:euaction}
We will give a simple calculation of the action of $\hek_\alpha$ on a gauge variant vertex. We refer to \cite{yang2017graphical} for more details about the graphical calculation method. Let us define
\begin{equation}
\langle g_e ||j,k\rangle:=\makeSymbol{\raisebox{0.\height}{\includegraphics[width=0.2\textwidth]{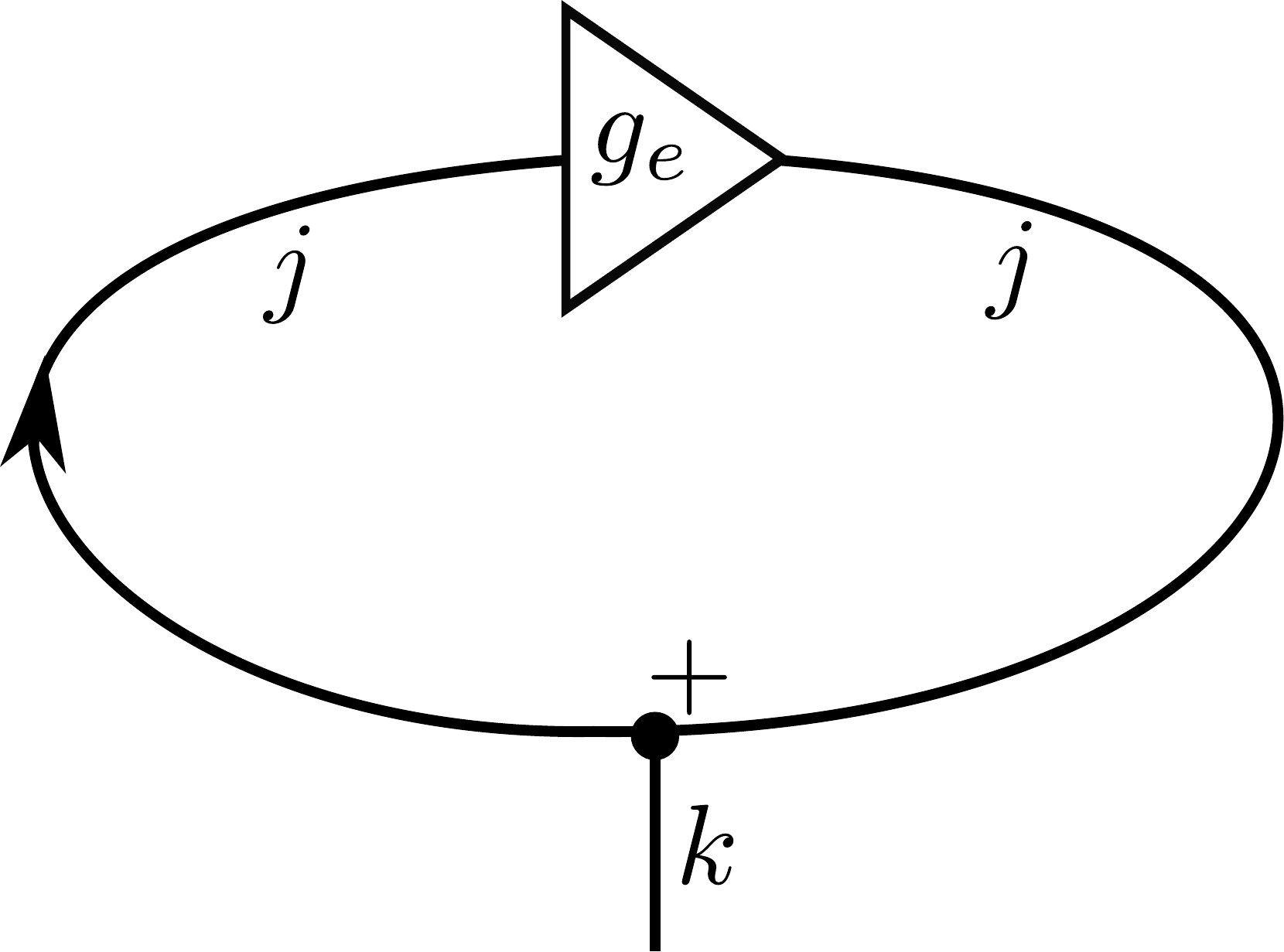}}}
\end{equation}
Then a straightforward calculation gives
\begin{equation}
\begin{aligned}
&\langle g_e|\hek_\alpha ||j,k\rangle=3\sqrt{6}\frac{1}{W_l}\sum_{J} d_JW_J^2\makeSymbol{\raisebox{0.\height}{\includegraphics[width=0.1\textwidth]{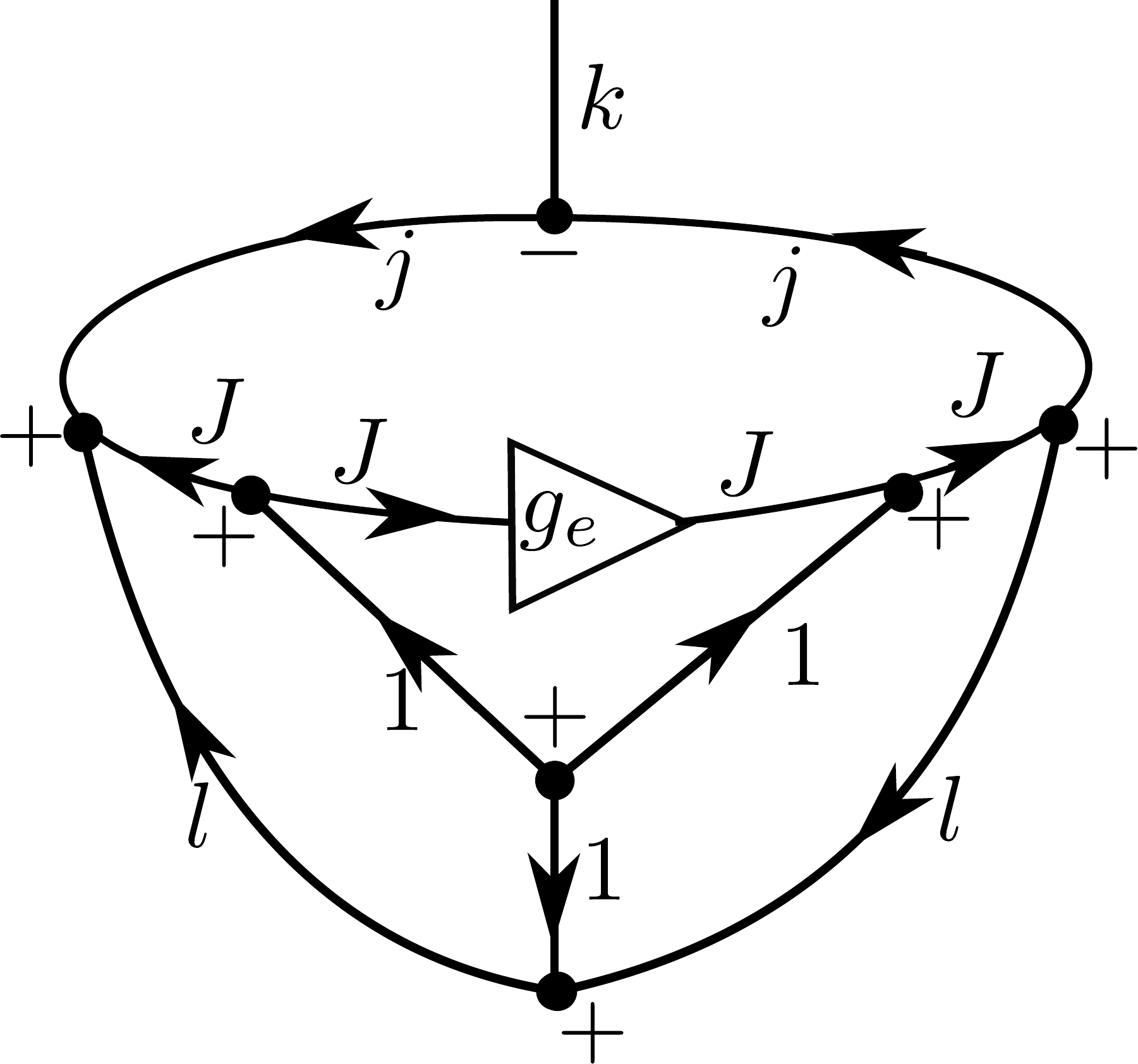}}}\\
=&3\sqrt{6}\frac{1}{W_l}\sum_Jd_JW_J^2\left(\sum_x d_x~\makeSymbol{\raisebox{0.\height}{\includegraphics[width=0.1\textwidth]{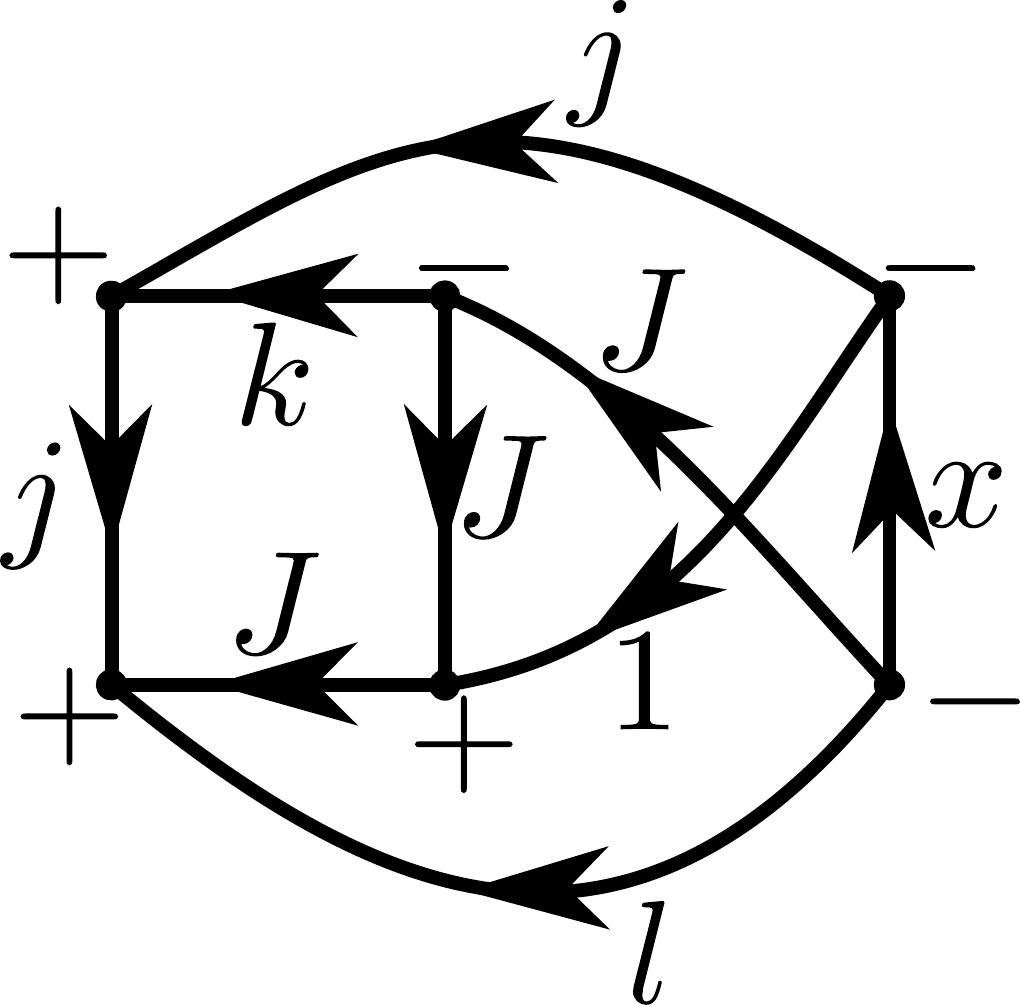}}}\makeSymbol{\raisebox{0.\height}{\includegraphics[width=0.1\textwidth]{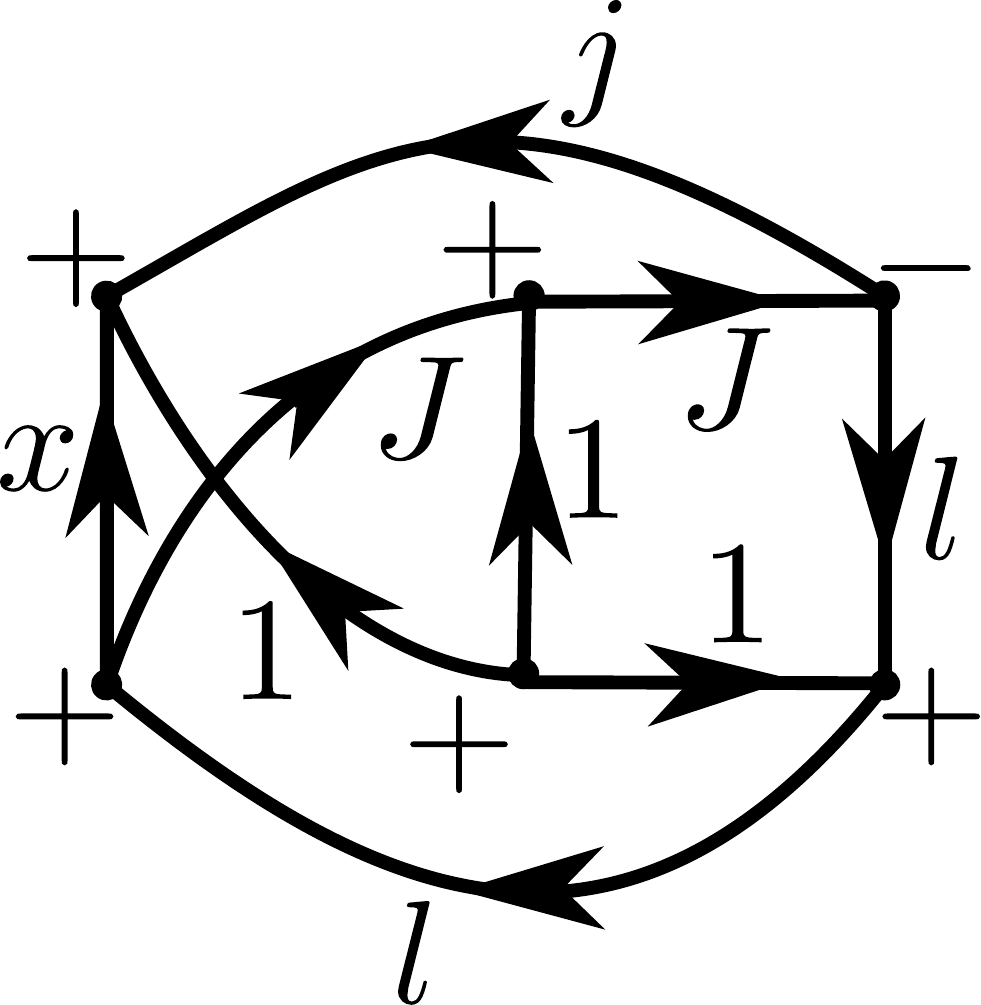}}}\right)\makeSymbol{\raisebox{0.\height}{\includegraphics[width=0.1\textwidth]{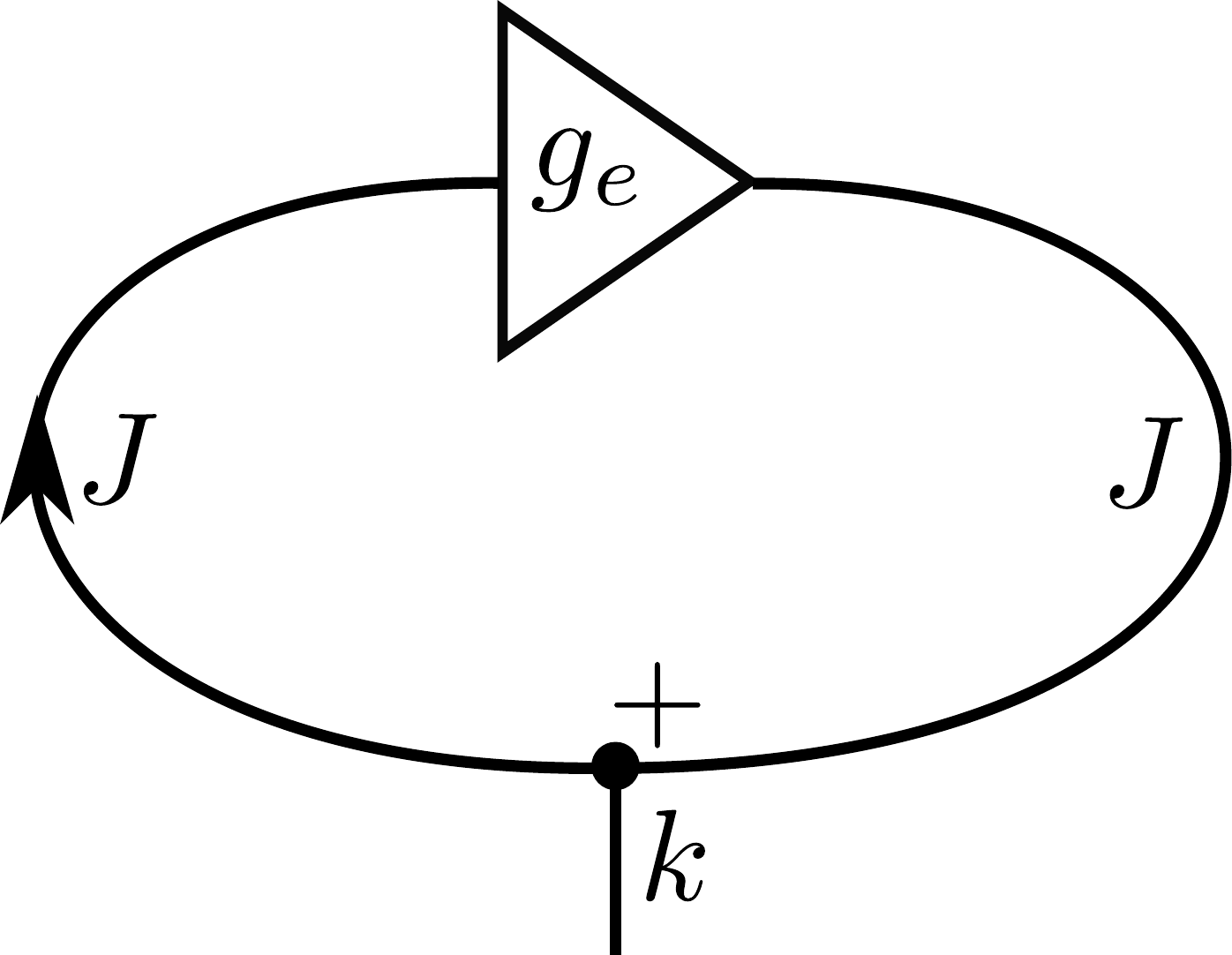}}}
\end{aligned}
\end{equation}
where a summation of two 9-$j$ symbols appears in the last line. Then we have
\begin{equation}
\begin{aligned}
&\hek||j,k\rangle\\
=&\frac{\sqrt{(2 j-k+1)(2j+k+2)}  (2 j+3+k (k+1))}{2 (2j+1) }||j+1/2,k\rangle\\
&-\frac{\sqrt{(2 j-k)(2 j+k+1)} (2 j-1-k (k+1))}{2(2 j+1)}||j-1/2,k\rangle.
\end{aligned}
\end{equation}
Because of $\langle j,k||j,k\rangle=\frac{1}{2j+1}$, we have
$|j,k\rangle=\sqrt{2j+1}||j,k\rangle$. Therefore, we can get
\begin{equation}
\begin{aligned}
&\hek|j,k\rangle\\
=&\frac{\sqrt{(2 j-k+1)(2j+k+2)}  (2 j+3+k (k+1))}{2 \sqrt{(2j+1)(2j+2)} }|j+1/2,k\rangle\\
&-\frac{\sqrt{(2 j-k)(2 j+k+1)} (2 j-1-k (k+1))}{2\sqrt{2j(2 j+1)}}|j-1/2,k\rangle.
\end{aligned}
\end{equation}

\section{A Weyl sequence of the operator $\hat{H}_I$}\label{app:weyl}
 Let $g:\mathbb{R}\to \mathbb{R}$ be a Schwartz function.
The Weyl sequence $\{|\omega\rangle_n\}$ of the operator $\hat{H}_I$  is defined by
\begin{equation}
|\omega\rangle_n:=\sum_{j}\frac{g(\frac{1}{n}\chi(j))}{\sqrt{n}}\varphi_\omega(j)|j\rangle.
\end{equation}
To prove that $|\omega\rangle_n$ really is a Weyl sequence, we will proceed in two steps.
 First we will show that
\begin{equation}
\lim_{n\to \infty}{_n}\langle \omega|\omega\rangle_n=4|A|^2\int_0^\infty |g(x)|^2\dd x.
\end{equation}
By choosing some large integer $j_0$ and introducing a notation $\mathcal{J}^+:=\{j_0,j_0+1,\cdots\}$ and $\mathcal{J}^-:=\{j_0+1/2,j_0+3/2,\cdots\}$, we have
\begin{equation}\label{eq:eq1}
\begin{aligned}
&{_n}\langle \omega|\omega\rangle_n=\frac{1}{n}\sum_j|g(\frac{\chi(j)}{n})|^2|\varphi_\omega(j)|^2\\
\leq&\frac{1}{n}\sum_{j=1/2}^{j_0-1/2}|g(\frac{\chi(j)}{n})|^2|\varphi_\omega(j)|^2+\frac{1}{n}\sum_{j\in \mathcal{J}^+}|g(\frac{\chi(j)}{n})|^2|f_\omega^+(j)-\varphi_\omega(j)|^2\\
&+\frac{1}{n}\sum_{j\in \mathcal{J}^-}|g(\frac{\chi(j)}{n})|^2|f_\omega^-(j)-\varphi_\omega(j)|^2+\frac{2}{n}\sum_{j\in \mathcal{J}^+}|g(\frac{\chi(j)}{n})|^2|f_\omega^+(j)-\varphi_\omega(j)||f_\omega^+|\\
&+\frac{2}{n}\sum_{j\in \mathcal{J}^-}|g(\frac{\chi(j)}{n})|^2|f_\omega^-(j)-\varphi_\omega(j)||f_\omega^-(j)|
\\&+\frac{1}{n}\sum_{j\in\mathcal{J}^+}|g(\frac{\chi(j)}{n})|^2 |f_\omega^+(j)|^2+\frac{1}{n}\sum_{j\in\mathcal{J}^-}|g(\frac{\chi(j)}{n})|^2 |f_\omega^-(j)|^2
\end{aligned}
\end{equation}
As in \eqref{eq:normeigen}, the terms apart from the summation of $|f_\omega^\pm(j)|^2$ are finite, and are denoted by $\psi_0$. Then we have
\begin{equation}\label{eq:normomegan}
\begin{aligned}
{_n}\langle \omega,k|\omega,k\rangle_n
\leq& \frac{ \psi_0}{n}+4|A|^2\sum_{j\in \mathcal{J}} |g(\frac{\chi(j)}{n})|^2\frac{\chi'(j)}{n} \\
&+\frac{4|A|^2}{n}\sum_{j\in\mathcal{J}^-}\Big\{|g(\frac{\chi(j)}{n})|^2  \chi'(j) \cos(2\omega\chi(j)+2\theta)\\
&-|g(\frac{\chi(j-1/2)}{n})|^2  \chi'(j-1/2) \cos(2\omega\chi(j-1/2)+2\theta)\Big\}
\end{aligned}
\end{equation}
where $A=:|A|e^{i\theta}$ and $\mathcal{J}=\mathcal{J}^+\cup\mathcal{J}^-$. Using the mean-value theorem, we obtain
\begin{equation}\label{eq:meanvalue}
\begin{aligned}
&\left| |g(\frac{\chi(j)}{n})|^2  \chi'(j) \cos(2\omega\chi(j)+2\theta)-|g(\frac{\chi(j-1/2)}{n})|^2  \chi'(j-1/2) \cos(2\omega\chi(j-1/2)+2\theta)\right|\\
=&\frac{1}{2}\left| \frac{\dd}{\dd x} \left(|g(\frac{\chi(x)}{n})|^2  \chi'(x) \cos(2\omega\chi(x)+2\theta)\right)\right|_{x=c(j)\in(j-1/2,j)}
\end{aligned}
\end{equation}
In the right hand side of \eqref{eq:meanvalue}, there will appear terms of multiplication of $\chi'(c(j))^2$ or $\chi''(c(j))$ with some bounded functions. Recalling that $\chi'(x)=1/ \xi(x+1/2)$, it can be obtained that both $\chi'(c(j))^2$ and $\chi''(c(j))$ decrease as $1/j^2$ as $j$ goes to infinity. Whence, there must exist some positive number $C$ such that
\begin{equation}\label{eq:boundedchi}
\frac{1}{2}\left| \frac{\dd}{\dd x}\left(|g(\frac{\chi(x)}{n})|^2  \chi'(x) \cos(2\omega\chi(x)+2\theta)\right)\right|_{x=c(j)\in(j-1/2,j)}\leq C\chi'(j)^2.
\end{equation}
Since $\int_{j_0}^\infty \chi'(x)^2\dd x<\infty$, we have $\sum_{j=j_0}^\infty\chi'(j)^2<\infty$. Equations \eqref{eq:normomegan}, \eqref{eq:meanvalue} and \eqref{eq:boundedchi} imply that
\begin{equation}
{_n}\langle\omega|\omega\rangle_n\leq \frac{\psi_1}{n}+4|A|^2\sum_{j\in \mathcal{J}} |g(\frac{\chi(j)}{n})|^2\frac{\chi'(j)}{n} 
\end{equation}
where all of the finite terms are absorbed into $\psi_1\in\mathbb{R}$. One can check easily that $\chi'(j)\leq \chi(j+1)-\chi(j)$. Thus we have
\begin{equation}
\begin{aligned}
{_n}\langle\omega|\omega\rangle_n\leq& \frac{ \psi_1}{n}+4|A|^2\sum_{j\in\mathcal{J}^+} |g(\frac{\chi(j)}{n})|^2\frac{\chi(j+1)-\chi(j)}{n}.
\end{aligned}
\end{equation}
Then we can get
\begin{equation}\label{eq:geq}
\begin{aligned}
&\lim_{n\to \infty}{_n}\langle\omega|\omega\rangle_n\leq \lim_{n\to \infty}4|A|^2\sum_{j\in\mathcal{J}^+} |g(\frac{\chi(j)}{n})|^2\frac{\chi(j+1)-\chi(j)}{n}=4|A|^2\int_0^\infty |g(y)|^2\dd y.
\end{aligned}
\end{equation}
Now if we replace the terms $|f_\omega^\pm(j)-\varphi_\omega(j)||f_\omega^\pm(j)|$ in \eqref{eq:eq1} by $-|f_\omega^\pm(j)-\varphi_\omega(j)||f_\omega^\pm(j)|$, $_n\langle \omega|\omega\rangle_n$ will be greater than or equal to the right hand side of  \eqref{eq:eq1}. Then using the fact that $\chi'(j)\geq \chi(j)-\chi(j-1)$, similar to the above analysis, we will obtain
\begin{equation}
\begin{aligned}
&\lim_{n\to\infty}{_n}\langle\omega|\omega\rangle_n\geq 4|A|^2 \int_0^\infty |g(y)|^2\dd y.
\end{aligned}
\end{equation}
Then we can get
\begin{equation}\label{eq:normalization}
\lim_{n\to \infty}\|~|\omega\rangle_n\|^2=4|A|^2 \int_0^\infty |g(y)|^2\dd y\leq \infty.
\end{equation}

The second step is to prove
\begin{equation}
\lim_{n\to\infty}(\omega-\hat{H}_I)|\omega\rangle_n=:\lim_{n\to \infty}L_\omega|\omega\rangle_n=0.
\end{equation}
One has
\begin{equation}
\begin{aligned}
&\|L_\omega|\omega\rangle_n\|^2\\
=&\frac{1}{n}\sum_j\Big|\Big(\xi(\sqrt{(j+1/2)(j+1)}~)\left[g(\frac{\chi(j+1/2)}{n})-g(\frac{\chi(j)}{n})\right]\varphi_\omega(j+1/2)\\
 &-\xi(\sqrt{j(j+1/2)}~)\left[g(\frac{\chi(j-1/2)}{n})-g(\frac{\chi(j)}{n})\right]\varphi_\omega(j-1/2)\Big)\Big|^2\\
\leq& \frac{\psi_0}{n}+\frac{1}{n}\sum_{j\in\mathcal{J}^+}\left|\Big(\sqrt{\xi(j+1/2)\xi(j+1)}\left[g(\frac{\chi(j+1/2)}{n})-g(\frac{\chi(j)}{n})\right]f_\omega^+(j+1/2)\right.\\
 &\left.-\sqrt{\xi(j)\xi(j+1/2)}\left[g(\frac{\chi(j-1/2)}{n})-g(\frac{\chi(j)}{n})\right]f_\omega^+(j-1/2)\Big)\right|^2\\
 &+\frac{1}{n}\sum_{j\in\mathcal{J}^-}\left|\Big(\sqrt{\xi(j+1/2)\xi(j+1)}\left[g(\frac{\chi(j+1/2)}{n})-g(\frac{\chi(j)}{n})\right]f_\omega^-(j+1/2)\right.\\
 &\left.-\sqrt{\xi(j)\xi(j+1/2)}\left[g(\frac{\chi(j-1/2)}{n})-g(\frac{\chi(j)}{n})\right]f_\omega^-(j-1/2)\Big)\right|^2.
\end{aligned}
\end{equation}
Substituting \eqref{eq:estimate_e}, we have
\begin{equation}
\begin{aligned}
&\|L_\omega|\omega\rangle_n\|^2\\
\leq& \frac{\psi_0}{n}+\frac{8|A|^2}{n}\sum_{j\in\mathcal{J}}\xi(j+1/2) \Big(\left[g(\frac{\chi(j+1/2)}{n})-g(\frac{\chi(j)}{n})\right]^2+ \left[g(\frac{\chi(j)}{n})-g(\frac{\chi(j-1/2)}{n})\right]^2\Big).
\end{aligned}
\end{equation}
 Finally, by using the mean-valued theorem again, we can obtain 
\begin{equation}
\begin{aligned}
&\|L_\omega|\omega\rangle_n\|^2
 \leq\frac{\psi_0}{n}+\frac{2|A|^2}{n}\sum_{j\in\mathcal{J}}\xi(j+1/2)\left(\left|g'(\frac{\chi(c_j^+)}{n})\frac{\chi'(c^+_j)}{n} \right|^2+ \left|g'(\frac{\chi(c_j^-)}{n})\frac{\chi'(c^-_j)}{n}\right|^2\right)
\end{aligned}
\end{equation}
with some numbers $c_j^+\in(j,j+1/2)$ and $c_j^-\in(j-1/2,j)$. 

By definition, $\xi(j+1/2)\chi'(c^\pm_j)=\frac{\chi'(c^\pm_j)}{\chi'(j)}$ and $g'(\frac{\chi(c_j^\pm)}{n})/g'(\frac{\chi(j)}{n})$ are bounded. Thus there exists some $C>0$ such that
\begin{equation}
\begin{aligned}
&\|L_\omega|\omega\rangle_n\|^2
 \leq\frac{\psi_0}{n}+\frac{C}{n^2}\sum_{j=j_0,j_0+1/2,j_0+1,\cdots} \left|g'(\frac{\chi(j)}{n})\right|^2\frac{\chi'(j)}{n}.
\end{aligned}
\end{equation}
Since $\sum_{j\in\mathcal{J}} \left|g'(\frac{\chi(j)}{n})\right|^2\frac{\chi'(j)}{n}\leq \infty$, we have finally
\begin{equation}
\lim_{n\to \infty}\|L_\omega|\omega\rangle_n\|^2\leq \lim_{n\to \infty} \frac{C}{n^2} \sum_{j\in\mathcal{J}} \left|g'(\frac{\chi(j)}{n})\right|^2\frac{\chi'(j)}{n} =0.
\end{equation}
This completes the proof to show that $|\omega\rangle_n$ is a Weyl sequence. 

\section{Normalization}\label{app:normalization}
We the normalization factor of states $|\omega\rangle$. 
As above, let $j_0$ be a large integer. Then we have
\begin{equation}\label{eq:ninethree}
\begin{aligned}
\langle\omega|\omega'\rangle=\psi_1+\sum_{j\in\mathcal{J}^+}\overline{f^+_{\omega}(j)}f^+_{\omega'}(j)+\sum_{j\in\mathcal{J}^-}\overline{f^-_{\omega}(j)}f^-_{\omega'}(j)\\
\end{aligned}
\end{equation}
where $\psi_1$ absorbs all the finite terms. Substituting $f^\pm_\omega(j)$ into \eqref{eq:ninethree} and using integration to estimate the summation, one has
\begin{equation}
\begin{aligned}
\langle\omega|\omega'\rangle
=&\psi_2+|A|^2\int _{j_0}^\infty \dd j\chi'(j)\left(e^{-i(\omega-\omega')\chi(j)}+e^{i(\omega-\omega')\chi(j)}\right)\\
&+|A|^2\int _{j_0+1/2}^\infty \dd j\chi'(j)\left(e^{-i(\omega-\omega')\chi(j)}+e^{i(\omega-\omega')\chi(j)}\right)\\
\end{aligned}
\end{equation}
where we reorganize all finite terms into $\psi_2$. Denoting $x:=\chi(j)$, $x_0:=\chi(j_0)$ and $x_1=\chi(j_0+1/2)$, we have
\begin{equation}
\begin{aligned}
\langle\omega|\omega'\rangle=&\psi_2+2|A|^2\int_{-\infty}^\infty\dd x e^{-i(\omega-\omega')x}-|A|^2\int_{-x_0}^{x_0}\dd x e^{-i(\omega-\omega')x}-|A|^2\int_{-x_1}^{x_1}\dd x e^{-i(\omega-\omega')x}\\
=&\psi_3+4\pi|A|^2\delta(\omega-\omega').
\end{aligned}
\end{equation}
According to section \ref{se:four}, $\langle \omega|\omega'\rangle$ must be proportional to the $\delta$-distribution. Thus we have $\psi_3=0$ and
\begin{equation}
\langle \omega|\omega'\rangle=4\pi|A|^2\delta(\omega-\omega').
\end{equation}
Therefore,the normalization of $\varphi_\omega(j)$ requires that $4\pi|A|^2=1$.

\section{Approximation}\label{app:saddle}
We provide a procedure to approximate the following integral 
\begin{equation}
I=\int_{-\infty}^\infty e^{\frac{1}{\sigma^2}\left(-\frac{x^2}{2}+i\sigma^2 f(x)\right)}\dd x=:\int_{-\infty}^\infty  e^{\frac{1}{\sigma^2}g(x)}\dd x.
\end{equation}
For $\sigma\ll 1$, $g(x)$ has a saddle point nearby $0$. Let the saddle point be 
\begin{equation*}
\mathring{x}=x_1\sigma+x_2\sigma^2+\cdots
\end{equation*}
Then we have
\begin{equation}
-(x_1\sigma+x_2\sigma^2+\cdots)+i\sigma^2 f'(x_1\sigma+x_2\sigma^2+\cdots)=0
\end{equation}
where $f(x)$ is real function. Expanding $f(x)$ around $x=0$, we obtain
\begin{equation}
0=-(x_1\sigma+x_2\sigma^2+\cdots)+i\sigma^2 f'(0)+i\sigma^2 f''(0)(x_1\sigma+x_2\sigma^2+\cdots)+\frac{i\sigma^2}{2}f'''(0)(x_1\sigma+x_2\sigma^2+\cdots)^2+\cdots
\end{equation}
which gives 
\begin{equation}
\begin{aligned}
x_1&=0\\
-x_2+if'(0)&=0\\
x_3&=0\\
-x_4 +if''(0)x_2+\frac{i}{2}f'''(0)(x_1)^2&=0\\
\cdots&\cdots
\end{aligned}
\end{equation}
Hence we have
\begin{equation}
\mathring{x}=if'(0)\sigma^2-f'(0)f''(0)\sigma^4+\cdots
\end{equation}
By the saddle point formula, we have
\begin{equation}
I= \frac{\sqrt{2\pi\sigma^2}}{\sqrt{-g''(\mathring{x})}}e^{\frac{g(\mathring{x})}{\sigma^2}}(1+O(\sigma^2))= \frac{\sqrt{2\pi\sigma^2}}{\sqrt{1-i\sigma^2f''(\mathring{x}) }}e^{ -\frac{\mathring{x}^2}{2\sigma^2}+if(\mathring{x})}(1+O(\sigma^2))
\end{equation}
where the branch of $\sqrt{-g''(\mathring{x})}$ is determined as $|\mathrm{Arg}(-g''(\mathring{x}))|<\pi/2$.

\begin{equation*}
f''(\mathring{x})=f''(0)+f'''(0)\mathring{x}+\frac{1}{2}f^{(4)}\mathring{x}^2+\cdots=f''(0)+O(\sigma^2)
\end{equation*}
\begin{equation*}
f(\mathring{x})=f(0)+f'(0)\mathring{x}+\cdots=f(0)+if'(0)^2\sigma^2+O(\sigma^4)
\end{equation*}
\begin{equation*}
\frac{\mathring{x}^2}{\sigma^2}=-f'(0)^2\sigma^2+O(\sigma^4)
\end{equation*}
Finally we have
\begin{equation}
I=\frac{\sqrt{2\pi\sigma^2}}{\sqrt{1-i\sigma^2f''(0) }}e^{ -\frac{f'(0)^2\sigma^2}{2}+if(0) }(1+O(\sigma^2))(1+O(\sigma^2)).
\end{equation}
In \eqref{eq:basicint}, we have
\begin{equation}
f(x)=\left(\phi-\frac{t}{2\sqrt{\omega_0}}\right) x+s\chi(j) x+s\theta(x+\omega_0)+s \chi(j)\omega_0
\end{equation}
Thus, we get
\begin{equation}
\begin{aligned}
f(0)&=s\theta(\omega_0)+s\chi(j)\omega_0\\
f'(0)&=\left(\phi-\frac{t}{2\sqrt{\omega_0}}\right) +s\chi(j) +s\theta'(\omega_0)\\
f''(0)&=s\theta''(\omega_0)
\end{aligned}
\end{equation}
which gives
\begin{equation}
I_j=\frac{\sqrt{2\pi\sigma^2}}{\sqrt{1-is\sigma^2\theta''(\omega_0) }}\exp\{ -\frac{\sigma^2\left(\phi-\frac{t}{2\sqrt{\omega_0}}+s\chi(j) +s\theta'(\omega_0)\right)^2}{2}+is\left(\theta(\omega_0)+\chi(j)\omega_0\right) \}.
\end{equation}

\end{document}